\documentclass[preprint,aps,amsmath,amssymb,showpacs,superscriptaddress,nofootinbib]{revtex4}
\usepackage{cancel} 
\usepackage{graphicx} 
\usepackage{array} 
\usepackage{makecell} 
\usepackage{xcolor}

\usepackage{hyperref}  
\hypersetup{ colorlinks=true, linkcolor=blue, citecolor=red, filecolor=magenta, urlcolor=cyan }

\newcommand{\Case}[2]{{\textstyle \frac{#1}{#2}}}

\begin{document}

\title{Consistent Truncation of Linearized gravitational waves in de Sitter space-time}

\author{Ghanashyam Date} \email{ghanashyamdate@gmail.com}

\affiliation{Chennai Mathematical Institute,\\ H1, SIPCOT IT Park, Siruseri, Kelambakkam 603
103 India}

\author{Harsh} \email{harsh22@cmi.ac.in}

\affiliation{Chennai Mathematical Institute,\\ H1, SIPCOT IT Park, Siruseri, Kelambakkam 603
103 India}

\begin{abstract}
	An important step in using observations of gravitational waves from bounded sources
	is to relate the observed waveforms far away from a source to the local dynamics and
	environment of the source. To isolate and identify different features of the source,
	multipole expansion of both the radiation field and the source distribution are
	crucial.  At a practical level, these expansions are truncated at some finite
	multipole order and a consistency issue arises.  This was flagged in \cite{CHK} as
	the emergence of disallowed $\log(r)$ terms in certain asymptotic forms and a
	particular quadrupolar truncation was proposed which was consistent with the source
	conservation.  

	In \cite{NKD80}, it was noted (incorrectly) that the particular solution does not
	satisfy the generalized harmonic gauge condition and this may be the source of
	appearance of the log(r) terms. In this work the consistency issue is analyzed by
	explicitly integrating the gauge conditions and the source conservation equations
	over the conformal time.  This gives a general procedure for a consistent
	truncation.  The procedure is verified in generalized harmonic gauge in conformal
	chart and in the Bondi gauge in Bondi chart. 
\end{abstract}


\maketitle

\tableofcontents 
\newpage

%
\section{Introduction} 
Changing background from Minkowski space-time to de Sitter space-time has a rather profound
effect on the analysis of weak gravitational waves, both conceptually and practically
\cite{ABK-I, ABK-II, ABK, ABK-III, NoIncoming, Saw1, Saw2, CFR, VRS, DH-1, CHK, NKD80, BBP}.
The practical computations also introduce pitfalls to watch out for.  One instance of this
is the computation of the approximated waveform in terms of the source truncated to finitely
many moments.  The retarded Green's function in the conformal chart and in generalized
harmonic gauge has a pre-factor such as $\Case{\eta}{R(\eta - R)}$ in some of the components
and a tail integral in another set of components.  By contrast, the corresponding Green's
function in Minkowski background has a factor $\Case{1}{R}$ for all components and there is
no tail integral.  These forms of the retarded Green's function in the de Sitter case
obstruct a closed form multipolar expansion of the particular solution.  A truncation of
such an expansion becomes necessary and an issue arises as to exactly how a truncation is to
be chosen.  In the recent work of Comp\`ere, Hoque and Kutluk \cite{CHK} the appearance of
log(r) in the Bondi-Sachs form of the solution was attributed to a source truncation being
inconsistent with the conservation equation.  They also proposed a definition of a
consistent quadrupolar truncation.  It was observed in \cite{NKD80} that the appearance of
disallowed log(r) is an indication that the truncated expression fails to be a solution of
the linearized equation i.e. fails to satisfy the wave equation and/or the gauge condition.
We elaborate on this issue and show how it can be addressed.  We show an alternative
consistent truncations for a solution satisfying only the generalized harmonic gauge
condition. This takes care of the mutual consistency of the wave equation, the gauge
condition and the conservation equation in the ``local'' or non-asymptotic region.  This
suffices to ensure that the solution also extends to the asymptotic region with the
admissible fall-off.

To articulate the issues raised by a truncation, as well as to clarify the computations, it
is useful to trace through the usual steps followed in the conformal chart (or the
cosmological chart with the conformal time $\eta$ replaced by the synchronous time $t$).
This is well known and is recalled for convenient reading in the section \ref{ConfChrt-sec}
below.  In the subsection \ref{EtaIntegration} we integrate the gauge condition and the
source conservation equation.  This is new and naturally suggests a general definition of
consistent truncation. This also highlights the role and the choice of boundary conditions
at the past cosmological horizon. In the following subsection \ref{TruncationConsistency} we
recall the definitions of source moments, discuss their conservation equations and present
the Comp\`ere et al truncation and an alternate truncation.  The forms of the conservation
equation and the generalized harmonic gauge condition, do {\em not} depend on any specific
solution. However, the particular solution does satisfy the gauge condition when the source
satisfies the conservation equation.  This is shown in appendix
\ref{GaugeConservationCheck}.  This is part one of the story.  

In the second part, we relate perturbations in the conformal chart and generalized harmonic
gauge to those in the Bondi-Sach chart in Bondi gauge.  This is done in the section
\ref{BSChrt-sec}.  This is a two step procedure. In the first step, the conformal chart
linearized solution is transformed by a coordinate transformation supplied by the background
de Sitter space-time. The transformed solution however does not have the Bondi-Sachs form of
the metric. In the second step, a suitable infinitesimal diffeomorphism is applied to bring
the transformed solution to the Bondi-Sachs form.  While these equations themselves do {\em
not} use any equations satisfied by the perturbations, we observe that the particular
solution we have already has a power series form as expected in the BS chart.  Therefore,
the differential equations determining the gauge generator are solved in a series form in
appendix \ref{GaugeGenerator}. The transformed metric coefficient $H_{AB}$ follows
straightforwardly and is also presented.  These steps do {\em not} depend on any truncation.
In the next appendix \ref{CalnD-AB}, the requirement of vanishing of $D_{AB}$ coefficient of
the asymptotic form in the BS chart is presented. This is valid for any multipole order and
is evaluated  for the quadrupolar solution given by Comp\`ere et al \cite{CHK}. It is also
evaluated for the solution truncated as per the order-2 alternate truncation defined in
\ref{TruncationConsistency}.  The last section includes a summary and conclusions. 
\section{Linearized equation and solution in Conformal chart} \label{ConfChrt-sec}
The relevant part of the de Sitter background space-time is the Poincare patch which is
conveniently described in the conformal chart. In this chart, the metric takes the form,
$\widehat{ds^2} = \Case{1}{H^2\eta^2}\big(-d\eta^2 + \vec{dx}\cdot\vec{dx}\big), \ H^2 :=
\Case{\Lambda}{3}$.  Denoting the de Sitter metric as $\hat{g}_{\mu\nu}(\eta,\vec{x})$, the
linearized metric is written as $g_{\mu\nu} = \hat{g}_{\mu\nu} + \epsilon h_{\mu\nu}$. In
terms of the trace reversed $\tilde{h}_{\mu\nu} := h_{\mu\nu} - \Case{h}{2}\hat{g}_{\mu\nu},
\ h := h_{\mu\nu}\hat{g}^{\mu\nu}$ and $B_{\mu} := \hat{\nabla}_{\alpha}
\tilde{h}^{\alpha}_{~\mu}$, the linearized Einstein equation becomes \cite{DH-1},
\begin{equation} \label{LinearisedEqn}
	\frac{1}{2}\Big[-\hat{\Box}\tilde{h}_{\mu\nu} + \big\{\hat{\nabla}_{\mu}B_{\nu} +
	\hat{\nabla}_{\nu}B_{\mu}
	-\hat{g}_{\mu\nu}\big(\hat{\nabla}^{\alpha}B_{\alpha}\big)\big\}\Big] +
	H^2\Big[\tilde{h}_{\mu\nu} - \tilde{h}\hat{g}_{\mu\nu}\Big] = 8\pi T_{\mu\nu} 
\end{equation}

The divergence of the left hand side is identically zero which implies the conservation of
the stress tensor as expected. The equation is invariant under the gauge transformations:
$\delta_{\xi}\tilde{h}_{\mu\nu} = \hat{\nabla}_{\mu}\xi_{\nu} + \hat{\nabla}_{\nu}\xi_{\mu}
- \hat{g}_{\mu\nu} \hat{\nabla}_{\alpha}\xi^{\alpha}$.

The equation is reduced to a simpler wave equation by choosing a gauge i.e. requiring that
$\tilde{h}_{\mu\nu}$ satisfy some additional conditions. Typically, one sets $B_{\mu} = 0$.
We refer to this choice as the {\em harmonic gauge}. In the resultant wave equation,
different components of $\tilde{h}_{\mu\nu}$ are coupled. A different choice, namely,
$B_{\mu} = 2H^2\eta\tilde{h}_{0\mu}$, referred to as the {\em generalized harmonic gauge}
\cite{VRS}, leads to decoupled wave equations which are conveniently solved. In terms of the
variables, $\chi_{\mu\nu} := H^2\eta^2\tilde{h}_{\mu\nu}$, the wave equations take the
decoupled form \cite{VRS},
\begin{equation}\label{WaveEqns}
	\Box\Case{\hat{\chi}}{\eta} =
	-\Case{16\pi\hat{T}}{\eta}\hspace{0.5cm},\hspace{0.5cm} \Box\Case{\chi_{0i}}{\eta} =
	-\Case{16\pi{T_{0i}}}{\eta}\hspace{0.5cm},\hspace{0.5cm} (\Box +
	\Case{2}{\eta^2})\Case{\chi_{ij}}{\eta} = -\Case{16\pi{T_{ij}}}{\eta}~.
\end{equation}
Here $\Box = -\partial^2_{\eta} + \partial^2_{i}$ is the d'Alembertian in Minkowski
space-time and $\hat{\chi} := \chi_{00} + \chi_i^{~i}$ and likewise $\hat{T}:= T_{00} +
T_i^{~i}$. The solutions of these decoupled wave equations {\em have to} satisfy the {\em
generalized harmonic gauge condition},
\begin{equation}\label{GaugeCondn}
	\partial^{\alpha}\chi_{\alpha\mu} + \frac{1}{\eta}\big(2\chi_{0\mu} +
	\delta^0_{\mu}\chi_{\alpha}^{~\alpha}\big) ~ = ~ 0 \ .
\end{equation}
Although the wave equation is decoupled, {\em this condition couples different components of
$\chi_{\mu\nu}$}. This is also true for the $\Lambda = 0$ case.

Both these equations are invariant under some residual gauge transformations \cite{DH-1},
\begin{eqnarray} 
	\delta\chi_{\mu\nu} & = & \big(\partial_{\mu}\underline{\xi}_{\nu} +
	\partial_{\nu}\underline{\xi}_{\mu} -
\eta_{\mu\nu}\partial^{\alpha}\underline{\xi}_{\alpha}\big) -
\frac{2}{\eta}\eta_{\mu\nu}\underline{\xi}_0 ~ ~ , ~ ~ \underline{\xi}_{\mu} :=
H^2\eta^2\xi_{\mu} = \eta_{\mu\nu}\xi^{\nu} ~ ~ , ~ ~ \mbox{with}\\
0 & = & \Box\underline{\xi}_{\mu} + \frac{2}{\eta}\partial_0\underline{\xi}_{\mu} -
\frac{2}{\eta^2}\delta^0_{\mu}\underline{\xi}_{0} \ .  \label{ResidualGauge}
\end{eqnarray}
Clearly, the $\delta\chi_{\mu\nu}$ satisfy the homogeneous wave equation.

The wave equations being inhomogeneous equations, their general solution is the sum of the
particular solution and a solution of the homogeneous equation: $\chi_{\mu\nu} :=
\bar{\chi}_{\mu\nu} + \underline{\chi}_{\mu\nu}$. The particular solution is determined by
the source and has the schematic form, $\bar{\chi}_{\mu\nu}(x) \sim \int d^4x'
G_{ret}(x,x')T_{\mu\nu}(x')$. The source being spatially compact, the integral ranges over
the source region while the solution is desired for $x$ in the far exterior region. The
gauge conditions (\ref{GaugeCondn}) are required to be satisfied by the {\em combined
solution} $\chi_{\mu\nu}$. Explicitly, the particular solution of the wave equation is given
by \cite{VRS,CHK},
\begin{eqnarray}\label{ParticularSoln}
	\bar{\hat{\chi}}(\eta,\vec{x}) & = & 4\int_{source}\frac{d^3x'}{R}
	\frac{\eta}{\eta-R} \hat{T}(\eta-R,\vec{x}') ~,~ R := |\vec{x}-\vec{x}'| ; \nonumber
	\\
	\bar{\chi}_{0i}(\eta,\vec{x}) & = & 4\int_{source}\frac{d^3x'}{R}
	\frac{\eta}{\eta-R} T_{0i}(\eta-R,\vec{x}') ~  ; \nonumber \\
	\bar{\chi}_{ij}(\eta,\vec{x}) & = & 4\int_{source}\frac{d^3x'}{R}
	T_{ij}(\eta-R,\vec{x}') ~+~ 4\int_{source}d^3x' \int_{-\infty}^{\eta-R}
	\frac{d\eta'}{\eta'}\partial_{\eta'} T_{ij}(\eta',\vec{x}')   ; \\
	& = & 4\int_{source}\frac{d^3x'}{R}\frac{\eta}{\eta - R} T_{ij}(\eta-R,\vec{x}') ~+~
	4\int_{source}d^3x' \int_{-\infty}^{\eta-R}
	\frac{d\eta'}{(\eta')^2}T_{ij}(\eta',\vec{x}')   . \nonumber
\end{eqnarray}
The last equation is an alternative form of the third equation, obtained by partial
integration.

Notice that shifting a solution by adding a residual gauge transformation
$(\delta\chi_{\mu\nu})_{res}$, is equivalent to adding a simultaneous solution of the
homogeneous equation and the generalized harmonic gauge condition. So if
$\bar{\chi}_{\mu\nu}$ satisfies the gauge condition, so must the
$\underline{\chi}_{\mu\nu}$.  It is a result stated in \cite{VRS} and verified in \cite{CHK}
that in the {\em exterior (source free) region}, the residual gauge invariance can be
exploited to set $\chi_{0i} = 0 = \hat{\chi}$. The gauge condition in equation
(\ref{GaugeCondn}), then implies that we can arrange to have $\chi_{00} = 0$ as well. With
these, we have $\chi_{0\mu} = 0$ which is referred to as the {\em synchronous gauge
condition}. The above specific residual gauge transformation taking a solution of the wave
equation and the generalized harmonic gauge condition to the synchronous gauge, guarantees
that the transformed spatial components satisfy $\partial^j\chi_{ji} = 0 = \chi_i^{~i}$.  We
refer to it as {\em spatial transverse, traceless} condition \cite{DH-1}.  With this gauge
fixing, it suffices to focus on the solution of the inhomogeneous wave equation for spatial
components alone together with the spatial $TT$ gauge condition.  It is important to keep in
mind that the {\em imposition of the synchronous gauge condition is done on the full
	solution and hence does not put any conditions on the stress tensor.}

{\em Note 1:} 

The residual gauge transformations are generated by those vector fields $\xi^{\mu}$ which
satisfy (\ref{ResidualGauge}). This is a differential specification and no boundary
conditions are stipulated.  In presence of source, the synchronous gauge can only be reached
in the exterior region. We may thus take these vector fields to vanish on the inner boundary
of the exterior region (which is  the same as the boundary of the compactly supported
source).  The other boundary of the exterior region is the future null infinity ($\eta =
0$), the past boundary of the Poincare patch ($\eta \to -\infty$) and the ``spatial
infinity'' ($|\vec{x}| \to \infty$ on each $\eta = $constant hypersurface).  All three
``intersect'' at the top right corner of the usual Penrose diagram of the de Sitter
space-time. Relevant for the gauge transformation which preserve the synchronous gauge, is
the behavior of $\xi^{\mu}$ at spatial infinity.  As noted in \cite{CHK}, these leftover
gauge transformations are of the form $\underline{\xi}_0 = -\Case{\eta}{3} \vec{\nabla}
\cdot\vec{D}\ , \ \underline{\xi}_i = D_i - \Case{\eta^2}{2} \nabla^2D_i$ where $\vec{D}$
depends only on $\vec{x}$ and satisfy $\nabla^2D_i = -\Case{1}{3} \partial_i
(\vec{\nabla}\cdot\vec{D}) \Rightarrow \nabla^2 (\vec{\nabla}\cdot\vec{D}) = 0$.  Thus
$\vec{\nabla}\cdot\vec{D}$ is harmonic in the exterior region and vanishes on the inner
boundary.  {\em If we take $\underline{\xi}_0$ to vanish at spatial infinity as well}, then
$\vec{\nabla}\cdot\vec{D} = 0$ everywhere in the exterior region.  This implies that $D_i$
for each $i$ is harmonic. {\em If $\underline{\xi}_i$ vanishes at spatial infinity} then
$D_i$ vanishes everywhere in the exterior and thus the leftover gauge transformation
vanishes everywhere in the exterior and we have a ``complete gauge fixing''.

One way to develop an approximation is to formulate a truncation procedure in terms of the
$(\chi_{ij})^{TT}$ which is ensured by the synchronous gauge. Hence the residual
transformations preserving the synchronous gauge also preserve the spatial TT condition,
even though $\chi_{ij}$ do change. The advantage is that it suffices to work with the
spatial components alone. This will be reported elsewhere. In this work we stay with the
generalized harmonic gauge. 

\subsection{Solving the gauge condition and conservation equation}\label{EtaIntegration}
Contrary to what was mentioned in \cite{NKD80}, it turns out that the particular solution
{\em does satisfy} the gauge condition! A direct verification, not using Fourier transform
as done in \cite{VRS}, is non-obvious which is what led to the erroneous statement.  This
check also led us to solve the gauge condition directly.

The gauge conditions (\ref{GaugeCondn}) are first order, {\em inhomogeneous} differential
equation in $\eta$ for $\hat{\chi}(\eta, \vec{x})$ and $\chi_{0i}(\eta, \vec{x})$ and can be
integrated as follows.

The $\mu = i$ gauge condition:
\begin{eqnarray}
	0 & = & \partial_0\chi_{0i} - \partial_j\chi_{ji} - \frac{2}{\eta}\chi_{0i} ,  ~ ~
	\mbox{Multiplying by $\eta^{-2}$} ~ \Rightarrow ~ \\
	& & \partial_0\left(\frac{\chi_{0i}}{\eta^2}\right) ~ = ~
	\frac{1}{\eta^2}\partial_j\chi_{ji} \nonumber \\
	\therefore \chi_{0i}(\eta, \vec{x}) & = & \eta^2\left[\int_{-\infty}^{\eta} d\eta'
	\frac{1}{\eta'^2}\partial_j\chi_{ji}(\eta',\vec{x}) + C_i(\vec{x})\right]; ~
	\mbox{which gives} ~ \label{Chi0iSoln} \\
& & \frac{1}{\eta}\partial_i\chi_{0i} ~ = ~ \eta\left[\int_{-\infty}^{\eta}d\eta'
\left\{\frac{1}{\eta'^2}\partial^{2}_{ij}\chi_{ij}(\eta', \vec{x})\right\} +
\partial_iC_i(\vec{x})\right] \ . \nonumber
\end{eqnarray}

The $\mu = 0$ gauge condition:
\begin{eqnarray}
	0 & = & \partial_0\hat{\chi} - \partial_{0}\chi_i^{\ i} - \partial_i\chi_{0i} -
	\frac{\hat{\chi}}{\eta} ~ ; ~~ \mbox{Multiplying by $\eta^{-1}$} \Rightarrow ~ ~ ~ ~
	\\
	\partial_0\left(\frac{\hat{\chi}}{\eta}\right) & = & \frac{1}{\eta}
	\big(\partial_0\chi_i^{\ i} + \partial_i\chi_{0i}\big) ~ = ~
	\frac{1}{\eta}\partial_0\chi_i^{\ i} + \eta\left\{\int_{-\infty}^{\eta}d\eta'
	\frac{1}{\eta'^2}\partial^2_{ij}\chi_{ij}(\eta', \vec{x})\right\} +
	\eta\partial_iC_i; \\
	\therefore \hat{\chi}(\eta, \vec{x}) & = & \eta\left[\int_{-\infty}^{\eta}d\eta'
	\frac{1}{\eta'}\partial_{\eta'}\chi_i^{\ i}(\eta',\vec{x}) +
\int_{-\infty}^{\eta}d\eta' \eta'\int_{-\infty}^{\eta'}d\eta'' \frac{1}{\eta''^2}
\partial^2_{ij}\chi_{ij}(\eta'', \vec{x}) + \hat{C}(\vec{x})\right]\ .  \label{ChiHatSoln}
\end{eqnarray}
In the last integration, the $\eta\partial_iC_i$ term from the middle equation gives
$\eta[\int_{-\infty}^{\eta}d\eta' \eta'](\partial_iC_i)$ which is divergent and hence we
require $\partial_iC_i = 0$ and drop this term.  The last equation includes the constant of
integration $\hat{C}(\vec{x})$.

Thus $\hat{\chi}$ and $\chi_{0i}$ admit the integration functions $\hat C(\vec x)$ and
$C_i(\vec x)$, with $\partial_iC_i=0$.  One must distinguish removable source-free
homogeneous additions from the contribution fixed by source boundary data.  The integration
function therefore separates as
\begin{equation} C_i=C_i^{\mathrm{free}}+C_i^{(\tau)},\qquad
\nabla^2C_i^{(\tau)}=-16\pi\tau_i .  \end{equation}
The source-free pieces solve the homogeneous wave equations only when they are harmonic and
may then be removed by a residual gauge transformation subject to the stated boundary
conditions.  By contrast, $C_i^{(\tau)}$ is harmonic only outside the source and must be
retained when the boundary angular-momentum mode is retained.  In what follows, only the
freely specifiable source-free parts of $C_i$ and $\hat C$ are set to zero.

The conservation equations for the stress tensor are \cite{DH-1}:
\begin{equation} \label{ConsEqn}
	-\partial_{\eta}T_{00} + \partial_iT_{0i} + \frac{\hat{T}}{\eta} ~ = ~ 0 ~ = ~
	-\partial_{\eta}T_{0i} + \partial_{j}T_{ji} + \frac{2}{\eta}T_{0i}\ .
\end{equation}
which are identical to the generalized harmonic gauge condition.  Hence these could also be
solved as,  
\begin{eqnarray}
	\hat{T}({\eta}, \vec{x}') & = & {\eta} \left[\int_{-\infty}^{{\eta}} d\eta'
		\frac{1}{\eta'}\partial_{\eta'}T_i^{\ i} (\eta',\vec{x}') +
		\int_{-\infty}^{{\eta}}d\eta' \eta' \int_{-\infty}^{\eta'} d\eta''
		\frac{1}{\eta''^2} \partial^2_{ij}T_{ij}(\eta'', \vec{x}') +
		\hat{\tau}(\vec{x}') \right]\ .  \label{THatSoln} \\
T_{0i}({\eta}, \vec{x}') & = & {\eta}^2\left[\int_{-\infty}^{{\eta}} d\eta'
\frac{1}{\eta'^2}\partial_j T_{ji}(\eta',\vec{x}') + \tau_i(\vec{x}')\right] ~ , ~
\partial_i \tau_i = 0 ; \label{T0iSoln}
\end{eqnarray}

We have exactly the same form of the constants of integrations which give divergent behavior
as ${\eta} \to -\infty$. These solutions are gauge invariant and to take care of these
constants, we have to stipulate boundary conditions. 

To begin with, in order to keep the source to be spatially compact we stipulate that
$\tau_i(x), \hat{\tau}(x)$ also have compact support and for convenience we take it to be
the same as that of the spatial components. If the spatial components are zero, then the
conservation equation (\ref{ConsEqn}) is a homogeneous equation for $T_{0i}$ whose solution
is given by $T_{0i}(\eta, \vec{x}) = \eta^2 \tau_i(\vec{x})$.  Integrating the $\hat{T}$
equation, requires $\partial_i\tau_i = 0$ and gives $\hat{T}(\eta, \vec{x}) =
\eta\hat{\tau}(\vec{x})$. Recall that the components of the stress tensor relative to a
locally inertial frame (in the de Sitter background) are given by $T_{inertial} = H^2\eta^2
T_{conformal}$ \cite{DH-1}.  We {\em assume $T_{ij} \sim \eta^{-2}$ as $\eta \to -\infty$ so
that the integrals are well defined}.  The above solutions then are dominated by the
$\hat{\tau}, \tau_i$ terms. If we demanded the $\hat{T}, T_{0i}$ also to behave as $T_{ij}$
then $\hat{\tau}, \tau_i$ have to be zero. A more general boundary condition is however
possible.

Although Poincare patch is not the full de Sitter space-time, it has been chosen as
appropriate and sufficient model for (classical) gravitational radiation from spatially
compact sources. Hence we have the three boundary components of future null infinity,
spatial infinity and the past cosmological horizon. The past cosmological horizon is an
interior boundary and no conformal compactification is involved. On this boundary, all null
congruences are shear free and thus have no incoming gravitational radiation.  There should
be no matter flux across the past horizon entering the Poincare patch.  The minimal
condition that we could impose is that any flux from a stress tensor can only be along the
horizon.  If $\ell^{\mu}, n^{\mu}, \ell^{\mu}n_{\mu} = -1$ are two future directed null
directions at any point on the horizon with $\ell^{\mu}$ being normal to the horizon, then we
require that $T_{\mu\nu}\ell^{\mu}\ell^{\nu} = 0$ \cite{NoIncoming}.  For $\ell^{\mu} \sim
(1, x^i/r)$, the condition becomes $T_{\mu\nu}\ell^{\mu}\ell^{\nu} \sim T_{00} + 2
T_{0i}\hat{x}^i + T_{ij}\hat{x}^i \hat{x}^j \to 0$ as $\eta \to -\infty$. As argued above,
with $T_{ij} \sim \eta^{-2}$ the leading contributions are $\hat{T} = \eta\hat{\tau}, T_{0i}
= \eta^2 \tau_i$. The boundary conditions as $\eta \to -\infty$ thus take the final form:
\begin{equation} \label{BdryCond}
	\hat{\tau} ~ = ~ 0~ , ~ \tau_i \hat{x}^i = 0 ~ , ~ \partial_i\tau_i ~ = ~ 0 ~ , ~
	T_{ij} \to \frac{\tau_{ij}(\vec{x})}{\eta^2} .
\end{equation}

As with the $\chi_{\mu\nu}$, {\em if} we are given $T_{ij}(\bar{\eta}, \vec{x}')$ which {\em
happens} to satisfy spatial $TT$ condition, then the remaining components determined as
above have only the contribution from the boundary terms.  More generally, if the truncation
of $\hat{T}, T_{0i}$ is deduced from a truncation of $T_{ij}$ using these solutions, the
conservation is guaranteed. The implications of the boundary conditions are discussed in the
next subsection.

When the inhomogeneous wave equation is used, the free variables in the solution of the
gauge conditions and those in the solution of the conservation equation get related. In
particular,  we get
\begin{equation}
	\nabla^2 C_i^{(\tau)} = -16\pi\tau_i ~ \mbox{and}~ C_i^{(\tau)}(\vec{x}) = 4\int
	d^3\vec{x}' \frac{\tau_i(\vec{x}')}{R} ~,~  \partial_i C_i^{(\tau)}(\vec{x}) = 0
	\Leftarrow \partial'_i\tau_i(\vec{x}') = 0\ .
\end{equation}

Here and throughout the paper, $C_i$ without a superscript denotes the complete integration
function appearing in (\ref{Chi0iSoln}), whereas $C_i^{(\tau)}$ denotes the source-induced
part fixed by the boundary datum $\tau_i$.  Thus $C_i=C_i^{\mathrm{free}}+C_i^{(\tau)}$;
after the removable source-free part is set to zero, all source-dependent formulae
consistently retain $C_i^{(\tau)}$.

A solution of the linearized Einstein equation can be constructed by {\em choosing}
$\chi_{ij}(\eta,\vec{x}) := \bar{\chi}_{ij}(\eta, \vec{x})$ given in (\ref{ParticularSoln})
and generating the remaining components from the above solutions of the gauge condition. The
so constructed components {\em do} match the $\bar{\hat{\chi}}, \bar{\chi}_{0i}$ given in
(\ref{ParticularSoln}) as can be verified with some algebra. This is verified in appendix
\ref{GaugeConservationCheck}.  Thus, {\em given} a $T_{ij}(\bar{\eta},\vec{x}')$ we can
generate a linearized wave. We are still free to add $(\delta\chi_{\mu\nu})_{res}$ to the
above construction. 

In summary, for both $\chi_{\mu\nu}$ and $T_{\mu\nu}$, from any given compactly supported
spatial components as functions of $(\eta, \vec{x})$ over the Poincare patch together with
the stipulated boundary behavior on the past horizon, the remaining components can be
constructed such that the gauge condition/conservation equations hold.  The only
restrictions are to ensure existence of the integrals and finiteness of the solutions.
These statements are valid for exact solutions or for truncated solutions of gauge condition
{\em provided} the truncations are generated from that of the spatial components.

\subsection{Truncation and a Consistency Issue} \label{TruncationConsistency}
%
%
A truncation refers to approximating a solution constructed from an approximated source. A
consistency issue arises because the (a) the source approximation may not be consistent with
the constraints on the source eg the conservation equation in our case and (b) the
constructed, approximate solution may not be consistent with the defining equation eg the
wave equation together with the gauge condition in our case. We will use the solutions of
the gauge condition and the conservation equation obtained in the previous section to ensure
that both consistencies are satisfied.  We begin with the source truncation.

A spatially compact source is conveniently described by its moments and the source may be
approximated by restricting the number of moments. A question arises as to how different
components of the stress tensor may be approximated. An obvious condition is to be
consistent with the source conservation equation. In the previous subsection we have already
established how consistency with conservation equation can be ensured.  We transcribe the
solution of the conservation equation (\ref{THatSoln},\ref{T0iSoln}) as well as the
conservation equation (\ref{ConsEqn}) in terms of the moments.  For convenience of reading,
we recall the definition of moments from \cite{CHK}.

Let $x_L := x_{i_1}\cdots x_{i_L}$ for $L \ge 0$ and let $a(\eta) := - \Case{1}{H\eta}$.
Define the {\em moments},
\begin{eqnarray} \label{MomentsDefns}
	Q_L(\eta) & := & \int_{\eta=Const}d^3x a^{L+1}(\eta)[T_{00}(\eta,\vec{x})] x_L
	\mbox{\hspace{0.9cm}(Mass moments)} \nonumber\\
	S_{|L}(\eta) & := & \int_{\eta=Const}d^3x
	a^{L+1}(\eta)[\delta^{ij}T_{ij}(\eta,\vec{x})] x_L \mbox{\hspace{0.5cm}(Pressure
	moments)} \nonumber\\
	P_{i|L}(\eta) & := & \int_{\eta=Const}d^3x a^{L+1}(\eta)[T_{0i}(\eta,\vec{x})] x_L
	\mbox{\hspace{1.0cm}(Current moments)} \nonumber\\
	S_{ij|L}(\eta) & := & \int_{\eta=Const}d^3x a^{L+1}(\eta)[T_{ij}(\eta,\vec{x})] x_L
	\mbox{\hspace{1.0cm}(Stress moments)} \\
	& \Rightarrow & S_{|L}(\eta) = \delta^{ij}S_{ij|L}(\eta) ~ ~ , ~ ~ \hat{Q}_L(\eta)
	:= Q_L(\eta) + S_{|L}(\eta).
\end{eqnarray}
All moments are automatically symmetric in the $i_1, \cdots, i_L$ indices. We will use the
$\hat{Q}_L$ moments. We have used $S_{|L}$ to denote the pressure moments since they are the
moments of the trace part of the stress tensor $T_{ij}$.

Multiply by $[a(\eta)]^{L+1} x_L$ the equations (\ref{THatSoln}, \ref{T0iSoln}) and
integrate over the source region.  Using the spatial compactness of the stress tensor, this
leads to the relation among the moments as, 
\begin{eqnarray}
	P_{i|i_1,\ldots,i_L}(a) & = & -L\frac{a^{L-1}}{H} \int_0^a da' \frac{1}{(a')^{L}}
	S_{i(i_1|i_2,\ldots,i_L)}(a') ~ + \frac{a^{L-1}}{H^2} \tau_{i|i_1,\ldots,i_L}
	\label{CurrentMomnts} \\
	\hat{Q}_{i_1,\ldots,i_L}(a) & = & a^L\int_0^ada'a'\frac{\partial ~}{\partial a'}
	\left(\frac{S_{|i_1,\ldots,i_L}}{(a')^{(L+1)}}\right) \nonumber \\
	& & ~ ~ ~ ~ + \frac{L(L-1)}{H^2}a^{L} \int_0^{a} \frac{da'} {(a')^{3}} \int_0^{a'}
	\frac{da''}{(a'')^{L-1}} S_{(i_1i_2|i_3,\ldots,i_L)} \label{QHatMomnts}
\end{eqnarray}

The synchronous time $t$ is defined through $\eta := - H^{-1}e^{-Ht} := -(Ha)^{-1}$.  So
$\partial_t = \Case{1}{a(t)}\partial_{\eta} = Ha\partial_a$. Evaluating $\partial_t$ as
$Ha\partial_a$ on the above expressions, the differential form follows as, 
\begin{eqnarray}\label{MomentConservEqn}
	\partial_tQ_L ~ = ~ H(LQ_L - S_{|L}) - L P_{(i_1|i_2\ldots i_L)} ~ & , & ~ 
	\partial_tP_{i|L} ~ = ~ (L-1)H P_{i|L} - L S_{i(i_1|i_2\ldots i_L)} 
\end{eqnarray}
These are the same as those given in \cite{CHK}. The boundary contributions $\tau_{i|L}$ are
included in the current moments.  These equations can of course be derived directly from the
(\ref{ConsEqn}) as is done usually.

{\em Consequences of the boundary condition:} Vanishing divergence of $\tau_i$ together with
the spatial compactness of the source implies, $\tau_{(i|i_1,\ldots,i_L)} = 0 \ \forall \ L
\ge 0$. In particular for $L=0$, the total momentum of the source, $P_{i|} = \tau_{i|} = 0$.
For $L=1$, the total angular momentum of the source, $J_i := \epsilon_{ijk}P_{j|k} =
\epsilon_{ijk}\tau_{j|k}$ can be non-zero. All the $\tau_{(i|i_1,\ldots,i_L)}$ are
determined entirely by the stress moments, independent of the moments of $\tau_i(\vec{x})$.
On the other hand, all current moments antisymmetrised as, $P_{[i|i_1],\ldots, i_L}$ are
independent of the stress moments and are given by
$H^{-2}a^{L-1}\tau_{[i|i_1,],\cdots,i_L}$.  Except for $L=1$, all these current moments {\em
diverge} at the future null infinity and hence we have to choose $\tau_{[i|i_1,],\cdots,i_L}
= 0\ \forall\ L \ge 2$.

Thus the boundary condition $(\partial_i\tau_i = 0)$ at the past cosmological horizon and
the regularity at future null infinity allows only $\tau_{[i|j]}$ to be non-zero. Since
there is no contribution from gravitational radiation at the past cosmological horizon, the
conserved Noether charges are the zero total linear momentum, possibly non-zero angular
momentum and the `energy' associated with the Killing vector $-H(\eta\partial_{\eta} +
x^i\partial_i)$. The boundary conditions ($\hat{\tau}(\vec{x}) = 0 =
\hat{x}^i\tau_i(\vec{x})$) on matter make this energy vanish.

The same conclusions can also be arrived at  from the conservation equation
(\ref{MomentConservEqn}). These equations for $Q_L, P_{i|L}$ are inhomogeneous, first order
equations with the stress moments serving as the source terms. Consider the current moments.
The homogeneous have exponential $t-$dependence. For $L \ge 2$, the solutions diverge at
null infinity. For $L = 0$ they would diverge at the past cosmological horizon. However, the
boundary condition puts this to zero. The $L=1$ is time independent. Thus $P_{i|j}$ are
constants. Again the symmetric part is killed by the boundary condition and the
antisymmetric part survives.  Consider the equation for the mass moments. Its homogeneous
solution again have exponential dependence and solution diverges at null infinity for all $L
\ge 1$. Hence, all these must also vanish, leaving only the constant zeroth moment. But this
again vanishes from the boundary condition in the past, $\hat{\tau} = 0$.  Hence, all
homogeneous solutions for the mass moments and all except the conserved antisymmetrised
first current moment vanish by the past boundary condition and regularity at future null
infinity. The non-trivial evolution of the moments arise from the stress moments which have
to be prescribed at the linearized order. The role of the boundary conditions is transparent
in the integrated conservation equation.

The approximation of the source can now be stipulated in terms of the stress moments alone
with the mass and current moments given in (\ref{QHatMomnts}, \ref{CurrentMomnts}) or by
using the conservation equation (\ref{MomentConservEqn}). 

Thus, truncations defined in terms of the stress moments $S_{ij|L}, S_{|L}$ will
automatically give the non-zero current and the $\hat{Q}$ moments which are a proxy for the
mass moments $Q_L$.  {\em Any additional condition put on the {\em symmetrized} $P_{I|L},
	\hat{Q}_L$, as done in the consistent truncation in \cite{CHK}, will translate into
restricting the stress moments i.e. conditions on the source itself}.  
\subsection{An alternate truncation} \label{AlternateTruncation}

We have disposed of the homogeneous solutions of the equations (\ref{MomentConservEqn}).
Apart from the conserved Noether charges, the mass and current moments are completely
determined by the $S_{ij|L}, S_{|L}$ moments. To see potentially non-zero moments in a
truncation, it suffices to look at the equations with the homogeneous terms dropped,
\begin{equation} \label{MomentConservInhomo}
	\partial_tP_{i|L} ~ = ~  - L S_{i(i_1|i_2\ldots i_L)} ~ ~ , ~ ~ 
	\partial_tQ_L ~ = ~ HS_{|L} - L P_{(i_1|i_2\ldots i_L)} \ .
\end{equation}
We define an {\em alternate source truncation of order $L_0$} by $S_{ij|L}=0$ for $L>L_0$.
The possibly non-zero moments are deduced from the above equation.  Here are some special
cases: 
\begin{enumerate}
	\item \underline{$L_0 = 0$} ($S_{ij}, {S_|}$ are non-zero): 

		The only possibly non-zero moments are: $P_{i|j},Q,Q_{ij}$;
		($P_i=0=\partial_tQ_i \Rightarrow Q_i=0$ seen earlier.)

	\item \underline{$L_0 = 1$} ($S_{ij}, {S_|}, S_{ij|k}, {S}_{|i}$ are non-zero):

		The only possibly non-zero moments are:
		$P_{i|j},P_{i|jk};Q,Q_i,Q_{ij},Q_{ijk}$.

	\item \underline{$L_0 = 2$} ($S_{ij}, S_|, S_{ij|k}, S_{|i}, S_{ij|kl}, {S}_{|ij}$
		are non-zero):

		The only possibly non-zero moments are: $P_{i|j},P_{i|jk},P_{i|jkl};
		Q,Q_i,Q_{ij},Q_{ijk},Q_{ijkl}$.
\end{enumerate}

Here is a comparison of the two truncations in a tabular form:
\begin{center}
	\begin{tabular}{|c|c|c|}
		\hline
		Truncation & Alternative & CHK \\
		\hline
		$L_0 = 0$ & $S_{ij}, S_{|}; P_{i|j};$ & $S_{ij}, S_{|}, P_{i};$\\
		& $ Q, Q_{ij}.$ & $Q.$\\
		\hline
		$L_0 = 1$ & $S_{ij}, S_{ij|k}, S_|, S_{|i};$ & $S_{ij}, S_{ij|k}, S_|,
		S_{|i}$ \\
		& $P_{i|j}, P_{i|jk};$ & $P_i, P_{i|j}; $ \\
		& $Q, Q_i, Q_{ij}, Q_{ijk}.$ & $Q, Q_i$. \\
		\hline
		$L_0 = 2$ & $S_{ij}, S_{ij|k}, S_{ij|kl}, S_|, S_{|k}, S_{|kl};$ & $S_{ij},
		S_{ij|k}, S_{ij|kl}, S_|, S_{|k}, S_{|kl}:$ \\
		& $P_{i|j}, P_{i|jk}, P_{i|jkl};$ & $P_i, P_{i|j}, P_{i|jk}; $ \\
		& $Q, Q_i, Q_{ij}, Q_{ijk}, Q_{ijkl}.$ & $Q, Q_i, Q_{ij}$. \\
		\hline
	\end{tabular}
\end{center}
Notice that in the quadrupolar CHK truncation, $P_{i|jkl}$ is forced to be zero, which
through conservation equation puts conditions on $S_{ij|kl}$ (equation 255 of \cite{CHK}).
Thus the consistent truncation defined by \cite{CHK} puts restrictions on the source through
the conservation equations\footnote{This may also be interpreted as implying that in the
quadrupolar truncation defined in \cite{CHK}, only the restricted stress tensor is
visible.}. 

{\em Note:} The tensor-harmonic analysis of de Sitter Teukolsky waves in
\cite{Harsh:2024kcl} and the recent octupolar solution \cite{Harsh:2026vmh} provides an
independent field-level confirmation that the CHK truncation is genuinely quadrupolar.  The
CHK solutions match the explicitly \(l=2\) electric- and magnetic-parity Teukolsky modes
written in Zerilli tensor harmonics.  Thus the quadrupolar character is manifest in the
multipole content of the gravitational field, as well as in the retained source moments.

Now we come to the construction of the correspondingly approximated particular solution.

We take the particular solution (\ref{ParticularSoln}) for the spatial components,
$\bar{\chi}_{ij}$. The integrand comprising the Green's function and the source $T_{ij}(\eta
-R, \vec{x}')$ are Taylor expanded about $\vec{x}' = 0$ on each $\eta = $constant
hypersurface.  This Taylor expansion is truncated to terms which allow only $S_{ij|L}$ for
$L \le L_0$. This is the same as the procedure discussed in \cite{CHK} for $\chi_{ij}$.  For
the remaining components of the particular solution, $\bar{\chi}_{0\nu}$, apply the same
procedure of Taylor expansion keeping terms up to the mass and current moments permitted by
the alternate truncation. The time derivatives of the mass and current moments are expressed
back in terms of the mass, current and stress moments using the conservation equations
(\ref{MomentConservEqn}). This defines our {\em truncation of $\bar{\chi}_{\mu\nu}(\eta,
\vec{x})$}.  In the appendix \ref{GaugeConservationCheck} it is shown that particular
solution with a conserved source, satisfies the generalized harmonic gauge condition at the
exact level. This also holds with the truncation. 

For illustration and contrast, we give the $\bar{\chi}_{\mu\nu}$ in the alternate truncation
at order-2, $L_0 = 2$.  

The $\bar{\chi}_{ij}$ expression is the same as in \cite{CHK}.  We display the expression
for $\bar{\chi}_{0i}$ and $\bar{\hat{\chi}}$ for comparison.  These are used in the appendix
\ref{CalnD-AB}. Not only there are more terms to order $\alpha^{5}$, the lower order terms
in this truncation are different from those in the Comp\`ere et al truncation. 
\begin{eqnarray} \label{Chi-ijSoln}
	\chi_{ij} & = & \chi_{ij}(-\infty) - 4 H^{2} P_{(i|j)} + 4 H^{2} n_{k} S_{ij|k} - 2
	H^{3} S_{ij|kk} + 2 H^{2} n_{k} n_{l}\, \partial_{t} S_{ij|kl} \nonumber \\
	& & + \alpha \Big( 4 S_{ij} + 4 n_{k}\, \partial_{t} S_{ij|k} - 2 H\, \partial_{t}
	S_{ij|kk} + 2 n_{k} n_{l}\, \partial_{t}^{2} S_{ij|kl} \Big) \nonumber \\
& & + \alpha^2 \Big( 4 n_{k} S_{ij|k} - 2 \partial_{t} S_{ij|kk} + 6 n_{k} n_{l}\,
\partial_{t} S_{ij|kl} \Big) + \alpha^3\Big( - 2 S_{ij|kk} + 6 n_{k} n_{l} S_{ij|kl} \Big)
\end{eqnarray} 
where $\alpha :=\left(\frac{1}{a(t_{ret})\rho}-H\right) = -H\left(\frac{\eta-\rho}{\rho}
+1\right) = -H\eta/\rho$ and the moments are functions of $(\eta -\rho)$. The $P_{(i|j)}$
term comes from the tail integral involving $S_{ij}$.

The $\chi_{0i}$ is obtained as, 
\begin{eqnarray} \label{Chi-0iSoln}
	\chi_{0i} & = &\Big( 4 P_i + 2 H^{2} P_{i\vert jj} - 2 H^{2} n_j n_k P_{i\vert jk} +
		4 n_j \partial_t P_{i\vert j} - 2 H\, \partial_t P_{i\vert jj} + 4 H^{2} n_j
		\partial_t P_{i\vert jkk} -\nonumber\\ 
	& & \tfrac{8}{3} H^{2} n_j n_k n_l \partial_t P_{i\vert jkl} + 2 n_j n_k
\partial_t^{2} P_{i\vert jk} - 2 H n_j \partial_t^{2} P_{i\vert jkk} + \tfrac{2}{3} n_j n_k
n_l\partial_t^{3} P_{i\vert jkl}\Big)\alpha \nonumber\\
& & + \Big(4 n_j P_{i\vert j} + 6 H^{2} n_j P_{i\vert jkk} - 6 H^{2} n_j n_k n_l P_{i\vert
	jkl} - 2 \partial_t P_{i\vert jj} + 6 n_j n_k \partial_t P_{i\vert jk} - 2 H n_j
	\partial_t P_{i\vert jkk} \nonumber\\
& & - 2 n_j \partial_t^{2} P_{i\vert jkk} + 4 n_j n_k n_l \partial_t^{2} P_{i\vert
jkl}\Big)\alpha^{2} + \Big(- 2 P_{i\vert jj} + 6 n_j n_k P_{i\vert jk} - 6 n_j \partial_t
	P_{i\vert jkk} \nonumber\\
& & + 10 n_j n_k n_l \partial_t P_{i\vert jkl}\Big)\alpha^{3} + \Big(- 6 n_j P_{i\vert jkk}
+ 10 n_j n_k n_l P_{i\vert jkl}\Big)\alpha^{4} 
\end{eqnarray} 
where $\hat{Q}_L=Q_L+S_{|L}$ , one can use the conservation equations in integral to replace
$Q_L$ and $P_{i|L}$ to write full $\chi_{\mu\nu}$ in terms of $S_{ij|L}$.

The vector solution has the same multipolar pattern as the scalar solution through the
powers $\alpha,\ldots,\alpha^4$.  More precisely, the coefficient of $\alpha^m$ in
$\chi_{0i}$ is obtained from $\hat\chi_m$ by the formal replacement
\begin{equation}
	\hat Q_L\longrightarrow P_{i|L}, \qquad m=1,\ldots,4,
	\label{ScalarToVectorReplacement}
\end{equation}
while keeping only the current moments with $L\leq3$.  Terms in $\hat\chi_m$ containing the
rank-four mass moment $\hat Q_{jklm}$ would require a rank-four current moment $P_{i|jklm}$
after this replacement.  However, such a moment is absent in our order-2 alternate
truncation and these terms are therefore dropped.  The coefficient $\hat\chi_5$ contains
only rank-four mass moments, which explains why $\chi_{0i}$ has no $\alpha^5$ term.

The scalar combination is
\begin{equation}\label{Chi-HatSoln}
	\hat{\chi}=\hat{\chi}_1\alpha+\hat{\chi}_2\alpha^2+
	\hat{\chi}_3\alpha^3+\hat{\chi}_4\alpha^4+\hat{\chi}_5\alpha^5 .
\end{equation}
The coefficient of $\alpha$ is
\begin{eqnarray*}
	\hat{\chi}_1 &=& 4\hat{Q}-2H^2n_in_j\hat{Q}_{ij}
	+\frac{3}{2}H^4\hat{Q}_{iikk}-3H^4n_in_j\hat{Q}_{ijkk}
	+\frac{3}{2}H^4n_in_jn_kn_l\hat{Q}_{ijkl} \nonumber\\
	&&+2H^2\hat{Q}_{ii}+4n_i\partial_t\hat{Q}_i
	-2H\partial_t\hat{Q}_{ii}+4H^2n_i\partial_t\hat{Q}_{ijj}
	-\frac{8}{3}H^2n_in_jn_k\partial_t\hat{Q}_{ijk} \nonumber\\
	&&-2H^3\partial_t\hat{Q}_{iikk} +H^3n_in_j\partial_t\hat{Q}_{ijkk}
	+2n_in_j\partial_t^2\hat{Q}_{ij} -2Hn_i\partial_t^2\hat{Q}_{ijj}
	+\frac{1}{2}H^2\partial_t^2\hat{Q}_{iikk} \nonumber\\
	&&+3H^2n_in_j\partial_t^2\hat{Q}_{ijkk}
	-\frac{5}{3}H^2n_in_jn_kn_l\partial_t^2\hat{Q}_{ijkl}
	+\frac{2}{3}n_in_jn_k\partial_t^3\hat{Q}_{ijk} -Hn_in_j\partial_t^3\hat{Q}_{ijkk}
	\nonumber\\
	&&+\frac{1}{6}n_in_jn_kn_l\partial_t^4\hat{Q}_{ijkl}.
\end{eqnarray*}
The coefficient of $\alpha^2$ is
\begin{eqnarray*}
	\hat{\chi}_2 &=& 4n_i\hat{Q}_i+6H^2n_i\hat{Q}_{ijj}
	-6H^2n_in_jn_k\hat{Q}_{ijk}-2\partial_t\hat{Q}_{ii} +6n_in_j\partial_t\hat{Q}_{ij}
	\nonumber\\
	&&-2Hn_i\partial_t\hat{Q}_{ijj} +13H^2n_in_j\partial_t\hat{Q}_{ijkk}
	-\frac{55}{6}H^2n_in_jn_kn_l\partial_t\hat{Q}_{ijkl}
	-\frac{5}{2}H^2\partial_t\hat{Q}_{iikk} \nonumber\\
	&&-2n_i\partial_t^2\hat{Q}_{ijj} +4n_in_jn_k\partial_t^2\hat{Q}_{ijk}
	+H\partial_t^2\hat{Q}_{iikk} -3Hn_in_j\partial_t^2\hat{Q}_{ijkk}
	-n_in_j\partial_t^3\hat{Q}_{ijkk} \nonumber\\
	&&+\frac{5}{3}n_in_jn_kn_l\partial_t^3\hat{Q}_{ijkl}.
\end{eqnarray*}
The coefficient of $\alpha^3$ is
\begin{eqnarray*}
	\hat{\chi}_3 &=& -2\hat{Q}_{ii}+6n_in_j\hat{Q}_{ij}
	-3H^2\hat{Q}_{iikk}+18H^2n_in_j\hat{Q}_{ijkk} -15H^2n_in_jn_kn_l\hat{Q}_{ijkl}
	\nonumber\\
	&&-6n_i\partial_t\hat{Q}_{ijj} +10n_in_jn_k\partial_t\hat{Q}_{ijk}
	+H\partial_t\hat{Q}_{iikk} -3Hn_in_j\partial_t\hat{Q}_{ijkk}
	+\frac{1}{2}\partial_t^2\hat{Q}_{iikk} \nonumber\\
	&&-6n_in_j\partial_t^2\hat{Q}_{ijkk}
	+\frac{15}{2}n_in_jn_kn_l\partial_t^2\hat{Q}_{ijkl}.
\end{eqnarray*}
The coefficient of $\alpha^4$ is
\begin{eqnarray*}
	\hat{\chi}_4 &=& -6n_i\hat{Q}_{ijj}+10n_in_jn_k\hat{Q}_{ijk}
	+\frac{3}{2}\partial_t\hat{Q}_{iikk} -15n_in_j\partial_t\hat{Q}_{ijkk}
	+\frac{35}{2}n_in_jn_kn_l\partial_t\hat{Q}_{ijkl}.
\end{eqnarray*}
Finally, the coefficient of $\alpha^5$ is
\begin{eqnarray*}
	\hat{\chi}_5 &=& \frac{3}{2}\hat{Q}_{iikk} -15n_in_j\hat{Q}_{ijkk}
	+\frac{35}{2}n_in_jn_kn_l\hat{Q}_{ijkl}.
\end{eqnarray*}
This concludes our illustration of the alternate truncation.

{\em Remark:} The truncation through spatial components as that we have described above,
aligns well with synchronous gauge fixing.  Exploiting the residual gauge freedom to go to
the synchronous gauge or equivalently taking the $TT$ part of the particular solution,
directly ensures that this truncation is consistent.  While this avenue of synchronous gauge
fixing is most natural within the context of linearized theory, it may not be suitable if
one desires to go to the non-linear ``post-de Sitter'' iterations.  In such a case, it is
better to keep the coordinates fixed during the iterations and may be go to the synchronous
gauge at the last stage. 

		
The expressions for $\chi_{ij}$, $\chi_{0i}$ and $\hat\chi$ in (\ref{Chi-ijSoln}),
(\ref{Chi-0iSoln}) and (\ref{Chi-HatSoln}) satisfy, to the retained order, the linearized
Einstein equation (\ref{LinearisedEqn}) as well as the generalized harmonic gauge condition
(\ref{GaugeCondn}), upon using the moment-conservation relations (\ref{MomentConservEqn}).
Here $P_i=0$ follows from the compact, divergence-free boundary datum $\partial_i\tau_i=0$
in (\ref{T0iSoln}).

While we have ensured that truncated solution satisfies the consistency conditions in the
local or non-asymptotic region, we have to verify whether it can represent radiation.  Now,
radiation per se is defined unambiguously only by an asymptotic stipulation. The most common
and almost canonical approach is to use the Bondi-Sachs form of the metric.  The full
non-linear Einstein equations are imposed on the BS form of the metric. For equations with
positive $\Lambda$, the admissible ansatz turns out to be a power series in $r$ with
coefficients related by the Einstein equations.  A truncated $\chi_{\mu\nu}$, should satisfy
these relations upon transforming to the BS gauge.  If the truncated $\chi_{\mu\nu}$
satisfies the linearized equations consistently in the non-asymptotic region, it may
potentially fail to satisfy the BS asymptotics and may put further restriction on the source
moments. This further potential violation does {\em not} occur for the solutions discussed
here.  

In the second part below, we go to the BS form and examine this aspect.
\section{To the Bondi-Sachs form of the solutions} \label{BSChrt-sec}
Bondi-Sachs approach is to choose coordinates which capture the idea of radiation escaping
from a bounded region (``isolated body'') to far away. The radiation is presumed to travel
along null geodesics and their spatial directions could be labeled by points on a sphere
surrounding the isolated body. The points along the geodesics themselves could be labeled in
relation to luminosity distance. This suggests coordinates $u,r,x^A, A = 1,2$. This presumed
interpretation is captured by the conditions: $g^{uu} = 0$ ($u=$ constant hypersurfaces are
null), $g^{uA} = 0$ (angular coordinates are constant along rays) and
$\partial_r\Case{\mbox{det}(g_{AB})}{r^4} = 0$ (the metric on the two dimensional surface
scale as $r^2$ i.e. $r$ is the luminosity distance).  These imply that the covariant
components of the metric satisfy $g_{rr} = 0 = g_{rA}$ \cite{CHK}.  These 3 conditions leave
$6$ free functions and the metric is conventionally taken in the form, now on referred to as
the BS form,
\begin{eqnarray} \label{BSFormEqn}
	ds^2_{BS} & = & \Case{V}{r}e^{2\beta} du^2 - 2e^{2\beta}dudr + g_{AB}(dx^A -
	U^Adu)(dx^{B} - U^Bdu) ~, ~ A, B = 1,2 ; \\
	& = & \left(\Case{V}{r}e^{2\beta} + g_{AB}U^AU^B\right) du^2 - 2e^{2\beta}dudr - 2
	\left(g_{AB}U^B\right) du\,dx^{A} + g_{AB}dx^Adx^B  ~, \\
	& := & g_{uu}du^2 + 2g_{ur}du dr +2g_{uA}du\,dx^A +g_{AB}dx^Adx^B \ .
\end{eqnarray}
The functions $V,\beta,g_{AB}, U^A$ are all functions of all the coordinates: $u \in
\mathbb{R}, r > 0$ with $x^A$ being coordinates on the sphere and may be taken to be the
standard $\theta, \phi$.  This is only a kinematic choice of coordinates adapted to the
expected interpretation\footnote{There is a coordinate freedom which preserves this BS form.
Fixing this freedom refers to Bondi gauge.} and the dynamics of Einstein equations links the
various metric components. In particular, the radiative fields are a {\em subclass} of
solutions of source free Einstein equations with a positive cosmological constant $\Lambda$.
For instance, the exact de Sitter solution for $\Lambda > 0$, is non-radiative and
corresponds to, 
\begin{eqnarray}\label{dSinBSEqn}
	\mbox{Choose} & : & \beta = 0 ~,~ \frac{V}{r} = -(1 -H^2r^2) ~,~ U^A = 0 ~,~ g_{AB}
	= r^2 \gamma_{AB}~, \\
	& & \mbox{where $\gamma_{AB}$ is the unit sphere metric.} \nonumber \\
	\overline{ds}^2_{de Sitter} & = & -(1-H^2r^2)du^2 - 2du dr + r^2(d\theta^2 +
	sin^2\theta d\phi^2) ~ \mbox{(BS coordinates)} ; \\
	& \leftrightarrow & g^{ur} = -1 ~,~ g^{rr} = 1 - H^2r^2 ~ , ~ g^{\theta\theta} =
	(g_{\theta\theta})^{-1} ~,~ g^{\phi\phi} = (g_{\phi\phi})^{-1}.
\end{eqnarray}
The radiative solutions are sought within the class of solutions deviating from the de
Sitter solution with certain stipulated asymptotic behavior as $r \to \infty$. 

It has been established that an ansatz with an $r^{-1}$ expansion with coefficients being
functions of $(u, \theta, \phi)$ is admissible.  In particular, \cite{CFR} shows that an
expansion of the form $g_{AB} = r^2q_{AB} + r C_{AB} + \cdots $ leads to the most general
solution of vacuum Einstein equation with $\Lambda > 0$ {\em without} generating any
$\log(r)$ terms.  Some of the metric coefficients have non-zero boundary values
\cite{CFR,BBP} which can be eliminated by suitable choice of coordinates on ${\cal I}$.
Different such boundary gauge choices have been made in \cite{CHK,BBP}.  We do not display
these here because we confine ourselves to expressing the conformal chart solution in the BS
chart.  We do use one additional condition to conform to \cite{CFR}, namely $g_{AB} = r^2
q_{AB}$ which automatically solves $\partial_r(det(g_{AB})/r^4) = 0$. The $g_{rr} = g_{rA} =
0$ together with this form of $g_{AB}$ is loosely referred to as {\em Bondi gauge}.  The
conditions $g_{rr}=g_{rA}=0$ together with the fixed determinant condition are referred to
here as {\em Bondi gauge}. 

This re-expressing of the solution in BS chart is a two step procedure: first apply the
coordinate transformation given in (\ref{BSTransformEqn}) to bring the background metric in
the BS form followed by an infinitesimal gauge transformation to bring the perturbed metric
in the BS form (\ref{BSFormEqn}). 

Using the usual definitions of $x^i(\rho, \theta, \phi)$, the de Sitter metric in conformal
coordinates can be expressed as
\begin{equation} \label{dSConfMetric}
	\overline{ds}^2_{dS} ~ = ~ \frac{1}{H^2\eta^2}\big(-d\eta^2 + d\rho^2 +
	\rho^2(d\theta^2 + sin^2\theta d\phi^2)\big) ~ .
\end{equation}
Let $x^A$ denote the spherical polar angles ($\theta, \phi$) in both the conformal and the
BS chart with coordinates, $(\eta, \rho, x^A)$ and $(u, r, x^A)$ respectively.  We have used
the same $A$ in both charts to emphasize that the angles are the same. The forms of the de
Sitter metric in the two charts, lead to the coordinate relations \cite{CHK}:
\begin{eqnarray} \label{BSTransformEqn}
	\eta(u,r) = -\frac{e^{-Hu}}{H(1+Hr)} & , &  \rho(u,r) = \frac{r e^{-Hu}}{1+Hr} ~ ~
	\leftrightarrow \\
	u(\eta,\rho) = -H^{-1}\ln\big(H\rho - H\eta\big) & , & r(\eta,\rho) =
	-\frac{\rho}{H\eta} \ .
\end{eqnarray}

Let us denote the conformal coordinate collectively by $\{x^{\mu}\}$ and the BS coordinates
by $\{\bar{x}^{\mu}\}$. The corresponding generic metrics are denoted by $g_{\mu\nu}$ and
$\bar{g}_{\mu\nu}$ respectively.  Given any metric in the conformal chart, it can be
transformed into the BS chart using the above transformation.  Subtracting the de Sitter
metric in BS chart, we infer the perturbations in the BS chart: $\epsilon \bar{h}_{\mu\nu}
:= \bar{g}_{\mu\nu} - \bar{g}^{dS}_{\mu\nu}$ which are the transforms of the perturbations
in the conformal chart: $\epsilon h_{\mu\nu} = g_{\mu\nu} - g_{\mu\nu}^{dS}$. They are
related as, 
\begin{eqnarray}
	\bar{g}_{\mu\nu}(\bar{x}) & := & \Case{\partial {x}^{\alpha}}{\partial
	\bar{x}^{\mu}} \Case{\partial {x}^{\beta}}{\partial \bar{x}^{\nu}}
	{g}_{\alpha\beta}(x(\bar{x})) ~ ~ \Rightarrow ~ ~ \\
	\bar{h}_{\mu\nu}(\bar{x}) & = & \Case{\partial \eta}{\partial\bar{x}^{\mu}}
	\Case{\partial \eta}{\partial\bar{x}^{\nu}} h_{00} + \big( \Case{\partial
		\eta}{\partial\bar{x}^{\mu}} \Case{\partial x^i}{\partial\bar{x}^{\nu}} +
		\Case{\partial x^i}{\partial\bar{x}^{\mu}} \Case{\partial
	\eta}{\partial\bar{x}^{\nu}} \big) h_{0i} + \Case{\partial
	x^i}{\partial\bar{x}^{\mu}} \Case{\partial x^j}{\partial\bar{x}^{\nu}} h_{ij} .
	\nonumber
\end{eqnarray}
The partial derivatives are easily worked out. The transformed $\bar{h}_{\mu\nu}$ are given
in appendix (\ref{BSGauge-sec}), equation (\ref{h-to-hbarTransform}). 

We denote the background de Sitter metric as $\hat{g}_{\mu\nu}$ when the chart is clear from
context.  We look for an infinitesimal vector field $\xi^{\mu} := \hat{g}^{\mu\nu}\xi_{\nu}$
such that $H_{\mu\nu} = \bar{h}_{\mu\nu} + \hat{\nabla}_{\mu}\xi_{\nu} +
\hat{\nabla}_{\nu}\xi_{\mu}$ satisfy the Bondi-Sachs form conditions: ${H}_{rr} = {H}_{rA} =
0$ and $\hat{g}^{AB}H_{AB} = 0$.  That is,
\begin{eqnarray} \label{BSGaugeCondEqn}
	H_{rr} = 0 & = & \bar{h}_{rr} + 2 \hat{\nabla}_{r}\xi_r ~ = ~ \bar{h}_{rr} +
	2\partial_r\xi_r - 2\hat{\Gamma}_{rr}^{\mu} \xi_{\mu} ~ ; \\
	H_{rA} = 0 & = & \bar{h}_{rA} + \hat{\nabla}_{r}\xi_A + \hat{\nabla}_A\xi_r ~ = ~
	\bar{h}_{rA} + \partial_r\xi_A + \partial_A\xi_r - 2\hat{\Gamma}_{rA}^{\mu}\xi_{\mu}
	~ ; \\
	\hat{g}^{AB}H_{AB} = 0 & = & \hat{g}^{AB}\big(\bar{h}_{AB} + \hat{\nabla}_A\xi_B +
	\hat{\nabla}_B\xi_A) ~ = ~ \hat{g}^{AB}\bar{h}_{AB} + 2\hat{\nabla}_A\xi^A \nonumber
	\\
	& = & \hat{g}^{AB}\bar{h}_{AB} + 2\big(\partial_A\xi^A + {\Gamma}^A_{AB}\xi^B +
	\hat{\Gamma}^A_{Aa}\xi^a\big) 
	~ = ~ \hat{g}^{AB}\bar{h}_{AB} + 2{D}_A\xi^A + \frac{4}{r}\xi^r
\end{eqnarray} 
The $\gamma_{AB}$ is the unit sphere metric and ${D}$ the corresponding covariant
derivative. 

Using the background connections $\hat{\Gamma}^{\mu}_{rr} = 0, \ \hat{\Gamma}_{rA}^{~B} =
\Case{\delta_A^B}{r}$,\ the Bondi-Sachs gauge conditions lead to,
\begin{eqnarray} \label{BSDifferentialCondEqn}
	& & \partial_r\xi_r = -\frac{1}{2}\bar{h}_{rr} ~ ~ , ~ ~ \partial_r\xi_A -
	\frac{2}{r}\xi_A = -\partial_A\xi_r -\bar{h}_{rA} ~ ~ , ~ ~ \xi^r =
	-\frac{1}{4r}\gamma^{AB}\bar{h}_{AB} - \frac{r}{2}{D}_A\xi^A 
\end{eqnarray}
The first two are inhomogeneous, first order differential equations in $r$ for $\xi_r (=
-\xi^{u}), \ \xi_A (= r^2\xi^A)$ while the last one gives $\xi^r$ in terms of the integrals
of the first two equations.  In turn, these determine 
\begin{equation}
	\xi_u ~ = ~ -(1-H^2r^2)\xi^u - \xi^r ~ = ~ 
	(1-H^2r^2)\xi_r + \frac{1}{4r}\gamma^{AB}\big(\bar{h}_{AB} + 2{D}_A\xi_B \big) .
	\label{Xi-u}
\end{equation}

These are the same as the equations given in \cite{CHK}. Notice that none of the quantities
appearing in these differential equations contain any explicit $\log(r)$. They can only
emerge from the solution of these, {\em inhomogeneous} differential equations.

The homogeneous solutions will of course give the residual gauge freedom of the partially
fixed Bondi gauges and these too have no $\log(r)$ terms.

The radial equation for $\xi_r$ shows that a log term can be generated in $\xi_r$ {\em
provided} $\bar{h}_{rr} \sim r^{-1}.$ The equations (\ref{BSDifferentialCondEqn}) are solved
in a series form in the appendix \ref{GaugeGenerator}.  There it is further established
that,
\begin{eqnarray}
	\xi_A(u, r,\theta,\phi) & = & r^2 \xi_A^{(-2)}(u,\theta,\phi) + r
	\xi_A^{(-1)}(u,\theta,\phi) + \sum_{k\ge 0}\frac{\xi_A^{(k)}(u,\theta,\phi)}{r^{k}}
	~ , ~ \mbox{with} \nonumber \\
	& & \xi_A^{(k)}(u,\theta,\phi)  = \left[ \frac{\partial_A
	\bar{h}_{rr}^{(k+2)}}{2(k+1)(k+2)} +\frac{\bar{h}_{rA}^{(k+1)}}{k+2} \right] ~
	\forall ~ k \ge 0 ~ ~ ; \\
\xi_r(u, r,\theta,\phi) & = & \xi_r^{(0)}(u,\theta,\phi) + \sum_{k\ge
1}\frac{\xi_r^{(k)}(u,\theta,\phi)}{r^{k}} ~ , ~ \mbox{with}  \nonumber \\
& & \xi_r^{(k)}(u,\theta,\phi)  = \frac{1}{2k}\bar{h}_{rr}^{(k+1)} ~ \forall~ k~ \ge 1  ~ ,
~  
\partial_A \xi_r^{(0)} ~ = ~ \xi_A^{(-1)} - \bar{h}_{rA}^{(0)} \ ; \\
\xi_u(u,r,\theta,\phi) & = & r^2(-H^2\xi_r^{(0)}) + r\left(-H^2\xi_r^{(1)}+ \frac{1}{4}
\gamma^{AB}\big( 2{D}_A\xi_B^{(-2)} + \bar{h}_{AB}^{(-2)} \big)\right) \nonumber \\ 
& & \hspace{0.4cm} + ~ \sum_{k\ge 0}\frac{1}{r^k}\left(\xi_r^{(k)} -H^2\xi_r^{(k+2)} +
\frac{1}{4} \gamma^{AB} \big( 2{D}_A\xi_B^{(k-1)} + \bar{h}_{AB}^{(k-1)} \big)\right) \ .
\end{eqnarray}
Here and below, coefficients outside their stated ranges are defined to vanish; in
particular, $\xi_r^{(j)}:=0$ for $j<0$.
The radial equations do not determine $\xi_A^{(-2)}$.  The equation
$\partial_A\xi_r^{(0)}=\xi_A^{(-1)}-\bar h_{rA}^{(0)}$ relates $\xi_A^{(-1)}$ to
$\xi_r^{(0)}$.  Thus the Bondi--Sachs conditions leave some freedom in the choice of
$\xi^\mu$.  In the appendix (\ref{GaugeGenerator}), the gauge transform of
$\bar{h}_{\mu\nu}$ to $H_{\mu\nu} = \bar{h}_{\mu\nu} + \hat{\nabla}_{\mu}\xi_{\nu} +
\hat{\nabla}_{\nu}\xi_{\mu}$ is computed. It takes the form,
\begin{eqnarray}
	H_{uu} := \sum_{k\ge -2}\frac{H_{uu}^{(k)}}{r^k} & ~ : ~ & H^{(k)}_{uu} = \bar
	h_{uu}^{(k)}+2\partial_u\xi_u^{(k)} -2H^2\xi_u^{(k+1)}+2H^2\xi_r^{(k+1)}
	-2H^4\xi_r^{(k+3)} ; \\
	H_{ur} := \sum_{k\ge 0}\frac{H_{ur}^{(k)}}{r^k} & ~ : ~ & H^{(k)}_{ur} =
	\bar{h}^{(k)}_{ur} + \partial_u\xi^{(k)}_r - (k-1)\xi_u^{(k-1)} + 2H^2\xi_r^{(k+1)}
	; \\
	H_{uA} := \sum_{k\ge -2}\frac{H_{uA}^{(k)}}{r^k} & ~ : ~ & H^{(k)}_{uA} =
	\bar{h}^{(k)}_{uA} +\partial_u\xi_A^{(k)} + \partial_A\xi_u^{(k)} \\
	H_{AB} := \sum_{k\ge -2}\frac{H_{AB}^{(k)}}{r^k} & ~ : ~ & H^{(k)}_{AB} =
	\bar{h}_{AB}^{(k)} + \big({D}_A\xi_B^{(k)} + {D}_B\xi_A^{(k)}\big) \nonumber \\
	& & \hspace{3.0cm} -\frac{1}{2}\gamma_{AB}\gamma^{CD}\big( \bar{h}_{CD}^{(k)} +
	2{D}_C\xi_D^{(k)}\big)
\end{eqnarray}
Together with the BS gauge conditions, this is the complete Bondi-Sachs form of the
linearized solution. The $H_{AB}$ is trace-free at each order and the coefficients
$\xi_A^{(-2)}, \xi_A^{(-1)}$ and $\xi_r^{(0)}$ signal residual gauge dependence. 

Below is the explicit identification of the BS form coefficients of the angular metric. The
$H_{AB}, \xi_A$ etc are explicitly dependent on the solutions $\chi_{\mu\nu}$. 
\begin{eqnarray}
	q_{AB} & = & \gamma_{AB} + H_{AB}^{(-2)} ~ \mbox{with} ~\\ 
	& & H_{AB}^{(-2)}  = \bar{h}_{AB}^{(-2)} + \big( {D}_A \xi_B^{(-2)} +
	{D}_{B}\xi_A^{(-2)}\big) - \frac{1}{2}\gamma_{AB}\gamma^{CD}\big(
	\bar{h}_{CD}^{(-2)} + 2 {D}_C \xi_D^{(-2)}\big) \nonumber \\
	C_{AB} & = & H_{AB}^{(-1)} ~ = ~ \bar{h}_{AB}^{(-1)} + \big( {D}_A \xi_B^{(-1)} +
	{D}_{B}\xi_A^{(-1)}\big) - \frac{1}{2}\gamma_{AB}\gamma^{CD}\big(
	\bar{h}_{CD}^{(-1)} + 2 {D}_C \xi_D^{(-1)}\big)  \\
	D_{AB} & = & H_{AB}^{(0)} ~ = ~ \bar{h}_{AB}^{(0)} + \big( {D}_A \xi_B^{(0)} +
	{D}_{B}\xi_A^{(0)}\big) - \frac{1}{2}\gamma_{AB}\gamma^{CD}\big( \bar{h}_{CD}^{(0)}
	+ 2 {D}_C \xi_D^{(0)}\big)  \\
	E_{AB} & = & H_{AB}^{(1)} ~ = ~ \bar{h}_{AB}^{(1)} + \big( {D}_A \xi_B^{(1)} +
	{D}_{B}\xi_A^{(1)}\big) - \frac{1}{2}\gamma_{AB}\gamma^{CD}\big( \bar{h}_{CD}^{(1)}
	+ 2 {D}_C \xi_D^{(1)}\big)  \ .
\end{eqnarray}

As per the general analysis \cite{CFR}, the $D_{AB}=0$ should hold. We verify this check for
the truncated solutions in appendix \ref{CalnD-AB}. 

In $\bar{h}$'s, we have the source moments evaluated at the retarded time $\bar{t}$ which is
defined through $\bar{\eta} := \eta - \rho =: -H^{-1}e^{-H\bar{t}}$. Their derivatives
w.r.t.  $\bar{t}$ evaluates as,
\begin{eqnarray}
	\partial_{\bar{t}} & = & \partial_{\bar{t}} \bar{\eta}\ \big(\partial_{\bar{\eta}}u\
	\partial_u + \partial_{\bar{\eta}}r\ \partial_r\big) = e^{-Hu}\big(e^{Hu}\partial_u
	+ 0\big) ~ = ~ \partial_u. 
\end{eqnarray}

We have all the ingredients to read off the coefficients of the asymptotic form. The only
input is the solution $\chi_{\mu\nu}$ expanded to any desired order.

An important consistency check is to verify $D_{AB} = 0$.  The calculation is done in the
appendix \ref{CalnD-AB}. Both the quadrupolar truncation of \cite{CHK} as well as the
alternate truncation at order-2 verify this condition. Other conditions can be verified
similarly.
\section{Summary}
Quite generally, a solution of the linearized Einstein equation is developed as a series and
for practical reasons a truncation is necessary. In the present context, a manageable form
of solution is obtained as a solution of an inhomogeneous wave equation satisfying certain
gauge condition with some residual gauge freedom still available.  There is an additional
issue if the constructed solution represents radiation from the presumed (conserved) source.
The radiative nature is naturally understood in the Bondi-Sachs form of the solution near
the null infinity. When a truncation is attempted two distinct consistency issues arise: (a)
the truncated solution may fail to be a solution of both the wave equations and the gauge
condition. If it satisfies both equations, at least we have a consistent approximate
solution in the ``local'' (non-asymptotic) region; (b) it may however fail to represent
``radiation'' i.e.  have the stipulated fall-off behavior. To test this we have to extend
and/or map the solution to the asymptotic region. The work of Compe\`re et al had found that
an inconsistent truncation done in the non-asymptotic region leads to inconsistent
asymptotic form (log(r) terms).  In the first part of the analysis, we constructed a
truncation consistent in the local region from a truncation of the spatial components of the
solution alone. We labeled this truncation by the highest stress moment $S_{ij|L}$ kept
calling it order $L$ truncation.  This approach also brought out the need and the role of
stipulating boundary behavior near the past boundary of the Poincare patch.  

That a truncation may fail the asymptotic consistency even if satisfies the local
consistency may be seen as follows.  Suppose the stress tensor has the form $T_{ij}(\eta,
\vec{x}) = Q_{ij}(\eta) g(\vec{x})$ where $g(\vec{x})$ has compact support.  Now use the
integral formulae (\ref{THatSoln}, \ref{T0iSoln}) to get $T_{0\mu}$. Source conservation
holds.  Also using the retarded solution local consistency is satisfied.  A possible source
of non-analyticity of the solution is the tail integral in $\chi_{ij}$. For a {\em choice}
$Q_{ij}(\eta') \sim \eta'$ the tail integral acquires a $\mathrm{log}(\eta -R) \sim
\log(\eta -\rho)$ dependence.  Near null infinity, $\eta \to 0_-$ the tail term acquires a
logarithmic dependence on $\rho$.  However this is not the way the null infinity is
approached. It is approached in terms of luminosity distance $r \to \infty$.  And $\eta -
\rho = -e^{-Hu}/H$ which is independent of $r$. Thus there is no $\log(r)$ type term from
the tail integral.  Additionally, it is also true that the $Q_{ij}(\eta) \sim \eta$ is
disallowed by the boundary condition at the past cosmological horizon which requires $T_{ij}
\to \tau_{ij}(\vec{x})/\eta^2$ i.e.  $Q_{ij} \sim \eta^{-2}$.  Thus, the possible source of
generation of non-analyticity in $r$, the tail integral, does not violate the asymptotic
behavior and local consistency suffices. 
\section*{Acknowledgments} It is a pleasure to thank Amitabh Virmani for many discussions,
careful reading of the draft and discussions on the verification of the $D_{AB} = 0$
condition. GD would like to thank Jahanur Hoque for discussions on the link between the
inconsistent truncation of the source and appearance of the log(r) terms, and K G Arun for
discussions about the flat space calculations. Harsh would like to thank Omkar Shetye for
discussions. GD is partially supported by Infosys Fellowship.

\appendix

\section{Mutual consistency of gauge condition and source conservation}
\label{GaugeConservationCheck}

In the section \ref{ConfChrt-sec}, we have noted that the generalized harmonic gauge
condition (\ref{GaugeCondn}) and the source conservation equation (\ref{ConsEqn}) have
identical form and can be integrated to express the $(0i)$ and the hatted components in
terms of integrals of the spatial components, equations (\ref{Chi0iSoln}, \ref{ChiHatSoln},
\ref{THatSoln}, \ref{T0iSoln}) respectively. We also have the particular solution which
expresses the $\chi_{\mu\nu}$ components in terms of the corresponding components of the
stress tensor given in equation (\ref{ParticularSoln}). Below we verify that these seven
equations are mutually consistent. 

The strategy is to begin with $\chi_{ij}$ given in terms of $T_{ij}$ (\ref{ParticularSoln}),
and substitute this in the $\chi_{0i}$ (\ref{Chi0iSoln}) and $\hat{\chi}$
(\ref{ChiHatSoln}). This gives these components in terms of $T_{ij}$. We want to rewrite
these back in terms of $T_{0i}, \hat{T}$. For this we use the $T_{0i}$ (\ref{T0iSoln}) and
$\hat{T}$ (\ref{THatSoln}) expressions in terms of $T_{ij}$.  These should match the
solutions $\chi_{\mu\nu}$ in terms of the $T_{\mu\nu}$ (\ref{ParticularSoln}).  This is
verified below.

Using $\chi_{ij}$ from equation (\ref{ParticularSoln}), we obtain $
\partial_{\eta}\chi^i_{~i}, \partial_j\chi_{ij} $ and $ \partial^2_{ij}\chi_{ij}$.
\begin{eqnarray}
	\partial_{\eta}\chi^i_{~i}(\eta,\vec{x}) & = & ~ ~ \ 4\int d^3\vec{x}' \frac{\eta}
	{R(\eta-R)} \partial_{\eta} T^i_{~i} (\eta-R, \vec{x}') ~ , ~ R = |\vec{x}-\vec{x}'|
	; \\
	\partial_j\chi_{ij}(\eta, \vec{x}) & = & -4 \int d^3\vec{x}' \Big[ \big( \partial_j
	R \big) \Big\{\frac{T_{ij}(\eta-R, \vec{x}')}{R^2} + \frac{\eta}{R(\eta-R)}
\partial_{\eta} T_{ij} (\eta-R, \vec{x}') \Big\} \Big] ; \\
	\partial_{ij}^2 \chi_{ij}(\eta, \vec{x}) & = & -4 \int d^3\vec{x}' \Big[
		\big(\partial^2_{ij}R\big) \Big\{ \frac{T_{ij}(\eta-R, \vec{x}')}{R^2} +
		\frac{\eta}{R(\eta-R)} \partial_{\eta} T_{ij}(\eta-R, \vec{x}')\Big\}
		\nonumber \\
& & \hspace{1.0cm} + \big(\partial_iR\ \partial_j R\big)\Big\{ -\frac{2}{R^3}T_{ij}(\eta -
R), \vec{x}') - \frac{\eta}{R(\eta-R)}\partial^2_{\eta}T_{ij}(\eta-R, \vec{x}') \nonumber \\
	& & \hspace{1.0cm} + \frac{\partial_{\eta}T_{ij}(\eta-R,
	\vec{x}')}{R}\Big(\frac{1}{R} + \frac{\eta}{\eta(\eta-R)} - \frac{\eta}{(\eta-R)^2}
	\Big) \Big\} \Big] \ .
\end{eqnarray}

\underline{Check for $\chi_{0i}$}:

We write the $\chi_{0i}$ in terms of $\chi_{ij}$ using (\ref{Chi0iSoln}) and denote the
final form of this expressions as (R.H.S.)$_{I}$.  Likewise, the final simplified form
obtained from (\ref{ParticularSoln}) as (R.H.S.)$_{II}$.  

The boundary term in (\ref{T0iSoln}) must be retained. Define
\begin{equation} C_i^{(\tau)}(\vec{x}) :=4\int d^3\vec{x}'\,
\frac{\tau_i(\vec{x}')}{|\vec{x}-\vec{x}'|}, \qquad \partial_iC_i^{(\tau)}=0, \qquad
\nabla^2C_i^{(\tau)}=-16\pi\tau_i .  \label{CiTauDefn} \end{equation}
\begin{eqnarray}
	\chi_{0i}(\eta, \vec{x}) & = & \eta^2 \left[\int_{-\infty}^{\eta} \frac{ds}{s^2}
	\partial_j \chi_{ij}(s,\vec{x}) +C_i^{(\tau)}(\vec{x})\right] \\
	& = & -4 \eta^2\int_{-\infty}^{\eta} \frac{ds}{s^2}\Big[\int d^3\vec{x}'\
	\partial_jR\Big\{ \frac{T_{ij}(s-R,\vec{x}')}{R^2} + \frac{s}{R(s-R)}
\partial_sT_{ij}(s-R,\vec{x}')\Big\} \Big] \nonumber \\
	& & \hspace{0.0cm} +\eta^2C_i^{(\tau)}(\vec{x}) \nonumber \\
	%
	%
& & \hspace{2.0cm} \mbox{Interchange integration order and the shift $s \to s + R$}
\nonumber \\
& & \nonumber \\
\therefore \mbox{(R.H.S.)}_{I} & = & -4\eta^2\int d^3\vec{x}' \partial_jR
\int_{-\infty}^{\eta-R} ds\Big\{ \underbrace{ \frac{T_{ij}(s,\vec{x}')}
{R^2(s+R)^2}}_{\frac{I_1}{R^2}} + \underbrace{\frac{\partial_sT_{ij}(s, \vec{x}') }
{R(R+s)s}}_{\frac{I_2}{R}} \Big\} +\eta^2C_i^{(\tau)}(\vec{x}) ;
\end{eqnarray}
The two integrals combine and simplify the r.h.s. of $\chi_{0i}$ as,
\begin{eqnarray}
	I_2 & = & \frac{T_{ij}(s,\vec{x}')}{s(s+R)}\Big|^{\eta-R}_{-\infty} +
	\int_{-\infty}^{\eta-R} ds\,T_{ij}(s,\vec{x}')\Big\{ -\frac{1}{s^2(s+R)}
	-\frac{1}{s(s+R)^2}\Big\} \nonumber \\
	& = & \frac{T_{ij}(\eta-R, \vec{x}')}{\eta(\eta-R)} +
	\int_{-\infty}^{\eta-R}ds\,T_{ij}(s,\vec{x}') \frac{2s+R}{s^2(s+R)^2} \\
	\therefore \frac{I_1}{R^2} + \frac{I_2}{R} & = &
	\frac{1}{R^2}\int_{-\infty}^{\eta-R}ds
	\Big[\frac{T_{ij}(s,\vec{x}')}{(s+R)^2}\Big\{1 + \frac{R(R+2s)}{s^2}\Big\}\Big] +
	\frac{T_{ij}(\eta-R,\vec{x}')}{R\eta(\eta-R)} \nonumber \\ & = &
	\frac{1}{R^2}\int_{-\infty}^{\eta-R} ds\frac{T_{ij}(s,\vec{x}')}{s^2} +
	\frac{T_{ij}(\eta-R,\vec{x}')}{R\eta(\eta-R)} ; \\
	\therefore \mbox{(R.H.S.)}_{I}  & = & -4\eta^2 \int d^3\vec{x}'\partial_j R\left[
\frac{1}{R^2}\int_{-\infty}^{\eta-R} \frac{ds}{s^2} T_{ij}(s,\vec{x}') + \frac{1}{\eta R
(\eta-R)} T_{ij}(\eta-R, \vec{x}') \right] \nonumber \\ & & +\eta^2C_i^{(\tau)}(\vec{x}).  ~
~ ~ \label{Chi)iFromGaugeSoln}
\end{eqnarray}
Now we write the $\chi_{0i}$ from (\ref{ParticularSoln}) and use the (\ref{T0iSoln}).
\begin{eqnarray}
	\chi_{0i}(\eta,\vec{x})\Big|_{from (\ref{ParticularSoln})} & = & 4 \int d^3\vec{x}'
	\frac{\eta}{R(\eta-R)} T_{0i}(\eta-R, \vec{x}') ~ ~ \mbox{Using (\ref{T0iSoln}), }
	\nonumber \\
	& = & 4 \int d^3\vec{x}' \frac{\eta}{R(\eta-R)} (\eta-R)^2
	\left[\int_{-\infty}^{\eta-R} ds \frac{1}{s^2}\partial'_jT_{ij}(s,\vec{x}')
	+\tau_i(\vec{x}')\right] \ \nonumber \\
\therefore \mbox{(R.H.S.)}_{II} & = & 4 \int d^3\vec{x}' \frac{\eta(\eta-R)}{R}
\int_{-\infty}^{\eta-R} ds \frac{1}{s^2}\partial'_jT_{ij}(s,\vec{x}') +4\eta\int
d^3\vec{x}'\, \frac{\eta-R}{R}\tau_i(\vec{x}').  ~ ~ ~ ~ ~\label{Chi0iFromParticuarSoln}
\end{eqnarray}
We have to flip the $\partial_j'$. 
\begin{eqnarray}
	\frac{\eta(\eta-R)}{R} \int_{-\infty}^{\eta-R}\frac{ds}{s^2}\partial_j'
	T_{ij}(s,\vec{x}') & = & \partial_j' \Bigg[\frac{\eta(\eta-R)}{R}
	\int_{-\infty}^{\eta-R} \frac{ds}{s^2}T_{ij}(s,\vec{x}')\Bigg] \nonumber \\
	& & - \partial_j' \Big\{\frac{\eta(\eta-R)}{R}\Big\}\cdot \int_{-\infty}^{\eta-R}
	\frac{ds}{s^2}T_{ij}(s,\vec{x}') \nonumber \\
	& & + \frac{\eta(\eta-R)}{R} (\partial'_jR)
	\frac{T_{ij}(\eta-R,\vec{x}')}{(\eta-R)^2} .
\end{eqnarray}
The first term is a divergence and vanishes upon $\int d^3\vec{x}'$ thanks to the
compactness of the source. Now use $\partial_j'R = - \partial_jR$ and simplify the second
term as $\partial_j\Case{\eta(\eta-R)}{R} = -\Case{\eta^2}{R^2}\partial_jR$.  This leads to
the bulk contribution below.  For the boundary contribution, \(\tau_{i|\emptyset}=\int
d^3x'\,\tau_i=0\) gives
\begin{equation} 4\eta\int d^3\vec{x}'\frac{\eta-R}{R}\tau_i(\vec{x}')
=\eta^2C_i^{(\tau)}(\vec{x}), \end{equation}
where the spatially integrated boundary mode vanishes.  Thus,
\begin{eqnarray}
	\mbox{R.H.S.}_{II} & = & 4 \int d^3\vec{x}' \Bigg[-\frac{\eta^2}{R^2}\partial_jR
	\int_{-\infty}^{\eta-R} \frac{ds}{s^2} T_{ij}(s, \vec{x}') -
\frac{\eta}{R(\eta-R)}\partial_jR T_{ij}(\eta-R, \vec{x}')\Bigg] \nonumber \\
& & +\eta^2C_i^{(\tau)}(\vec{x}) \nonumber \\
& = & -4\eta^2 \int
d^3\vec{x}'\partial_jR\Bigg[\frac{1}{R^2}\int_{-\infty}^{\eta-R}\frac{ds}{s^2}
T_{ij}(s,\vec{x}') + \frac{1}{\eta R (\eta-R)} T_{ij}(\eta-R, \vec{x}') \Bigg] \nonumber \\
& & +\eta^2C_i^{(\tau)}(\vec{x})
\end{eqnarray}
The two right-hand sides agree, verifying the \(\chi_{0i}\) check.  Moreover,
\(\partial_iC_i^{(\tau)}=0\), so the source-induced boundary term does not alter the
subsequent \(\mu=0\) gauge-condition check for \(\hat\chi\).

\underline{Check for $\hat{\chi}$}:

Having verified the $\chi_{0i}$ solution, we can use that in the verification of the
$\hat{\chi}$ solution. An alternate form for the integrating the $\hat{\chi}$ and likewise
for $\hat{T}$ is more convenient. These follow from the $\mu=0$ gauge condition which can be
expressed as,
\begin{eqnarray}
	\partial_{\eta}\frac{\hat{\chi}}{\eta} & = &
	\frac{1}{\eta}\Big(\partial_{\eta}\chi^i_{~i} + \partial_i\chi_{0i}\Big) ~ ~
	\Rightarrow \\
	\frac{\hat{\chi}}{\eta} & = & \int_{-\infty}^{\eta}\frac{ds}{s}\big\{\partial_s
	\chi^i_{~i}(s,\vec{x}) + \partial_i \chi_{0i}(s, \vec{x})\big\} ; ~ ~  \mbox{And
	likewise,} \label{IntegratedChi0iSoln}\\
	\frac{\hat{T}}{\eta} & = & \int_{-\infty}^{\eta}\frac{ds}{s}\big\{\partial_s
	T^i_{~i}(s,\vec{x}) + \partial_iT_{0i}(s, \vec{x})\big\} . 
\end{eqnarray} 
We have integrated the $\hat{\chi}$ in terms of $\chi^i_{~i}$ and $\chi_{0i}$. We already
have $\partial_{\eta}\chi^i_{~i}$ and need $\partial_i \chi_{0i}$. For the latter, we can
directly use the (\ref{ParticularSoln}). This gives us, 
\begin{equation}
	\partial_i\chi_{0i} ~ = ~ 4 \int d^3\vec{x}'\partial_i\Big\{\frac{\eta}{R(\eta-R)}
	T_{0i}(\eta-R, \vec{x}')\Big\}.
\end{equation}
Consider the R.H.S. of the integrated solution (\ref{IntegratedChi0iSoln}) as,
\begin{equation*}
	\frac{\hat{\chi}(\eta, \vec{x})}{\eta}  =  4 \int_{-\infty}^{\eta} \frac{ds}{s} \int
	d^3\vec{x}' \Big[\underbrace{\frac{s}{R(s-R)}\partial_sT^i_{~i}(s-R,
	\vec{x}')}_{\textcircled{\scriptsize{1}}} +
\underbrace{\partial_i\Big\{\frac{s}{R(s-R)} T_{0i}(s-R,
\vec{x}')\Big\}}_{\textcircled{\scriptsize{2}}} \Big] .
\end{equation*}
The factor of $s$ cancels. Interchange the order of integration. In the first term, shifting
$s \to s + R$, gives,
\begin{equation*}
	\textcircled{\scriptsize{1}} = 4 \int d^3\vec{x}' \int_{-\infty}^{\eta-R}ds
	\frac{1}{sR}\partial_sT^i_{~i}(s,\vec{x}') \ .
\end{equation*} 
The integrand of the second term is manipulated as,
\begin{eqnarray}
	\textcircled{\scriptsize{2}} & = & \partial_i\Big( \frac{1}{R(s-R)}\Big) \cdot
	T_{0i}(s-R,\vec{x}') + \frac{1}{R(s-R)}\big(-\partial_i R \big)
	\partial_0(T_{0i}(s-R, \vec{x}')) \nonumber \\
	& = & \left[-\partial'_i\Big(\frac{1}{R(s-R)}\Big) T_{0i}(s-R,\vec{x}')\right] +
	\frac{1}{R(s-R)}\big(\partial'_i R \big) \partial_{0}(T_{0i}(s-R, \vec{x}')) 
\end{eqnarray}
Here $\partial_0$ refers to differentiating the first argument of $T_{0i}$. In the first
term, $\partial'_i$ can be flipped onto $T_{0i}$ under the spatial integral using the
compactness of the source. This $\partial'_i$ acts on both arguments of $T_{0i}$. In the
second term, $\partial_i'R \partial_{0}T_{0i}(s-R,\vec{x}') = -
\partial'_{i,1}T_{0i}(s-R,\vec{x}')$. Here $\partial'_{i,1}$ denotes differentiation of the
first argument only while $\partial'_{i,2}$ denotes action on the second argument. The sum
of these two thus gives $\partial'_{i,2}T_{0i}(s-R,\vec{x}')$. Now shift $s \to s-R$. Thus,
\begin{eqnarray}
	\textcircled{\scriptsize{2}} & = & 4\int d^3\vec{x}' \int_{-\infty}^{\eta-R} ds
	\frac{1}{R s} \partial'_{i,2}T_{0i}(s, \vec{x}') \ .
\end{eqnarray}
Combining the two, we get the integrated solution as,
\begin{eqnarray}
	\frac{\hat{\chi}(\eta,\vec{x})}{\eta} & = & 4 \int d^3\vec{x}' \Bigg[
	\int_{-\infty}^{\eta-R} ds \frac{1}{Rs}\partial_sT^i_{~i}(s,\vec{x}') +
\frac{1}{Rs}\partial'_i T_{0i}(s,\vec{x}') \Bigg] \\
& = & 4 \int d^3\vec{x}' \frac{1}{R} \frac{\hat{T}(\eta-R,\vec{x}')}{\eta-R} ~ ~ \mbox{or,}
\\
\hat{\chi}(\eta,\vec{x}) & = & 4 \int d^3\vec{x}' \frac{\eta}{R(\eta-R)} \hat{T}(\eta-R,
\vec{x}') .
\end{eqnarray}
The R.H.S. of the last equation is exactly the particular solution (\ref{ParticularSoln}).

This completes the proof of mutual consistency of the integrated gauge condition and
integrated source conservation.
\section{de Sitter in Bondi-Sachs chart} \label{BSGauge-sec}
The variable $\chi_{\mu\nu}=H^2\eta^2\tilde h_{\mu\nu}$ used in the main text is built from
the trace-reversed field.  The coordinate transformation below instead acts on the
perturbation $h_{\mu\nu}$. We therefore define the {\em scaled perturbation} $\psi_{\mu\nu}$
as,
\begin{equation} 
	\psi_{\mu\nu}:=H^2\eta^2h_{\mu\nu},\qquad \psi_{00}=\frac{\hat\chi}{2},\qquad
	\psi_{0i}=\chi_{0i},\qquad \psi_{ij}=\chi_{ij}+\delta_{ij}\left(\frac{\hat\chi}{2} -
	\chi_{aa}\right).
	\label{PhysicalPerturbationReconstruction} 
\end{equation}
Thus $\chi_{\mu\nu}$ denotes the trace-reversed solution, while $\psi_{\mu\nu}$ denotes the
scaled perturbation, not necessarily a solution, used in the coordinate transformation.

{\em From conformal chart $h_{\mu\nu}$ to the BS chart $\bar{h}_{\mu\nu}$:} 
\begin{eqnarray} \label{h-to-hbarTransform}
	\bar{h}_{uu} & = & \left[\frac{e^{-2Hu}}{(1+Hr)^2}\right] \big(h_{00} -2Hrn^ih_{0i}
	+ (Hr)^2n^in^jh_{ij}\big)  \\
	& = & \big(\psi_{00} - 2Hr\ n^i\psi_{0i} + (Hr)^2 n^in^j\psi_{ij}\big) \nonumber \\
	\bar{h}_{ur} & = & \left[\frac{e^{-2Hu}}{(1+Hr)^3}\right] \big(h_{00} +
	(1-Hr)n^ih_{0i} -Hrn^in^jh_{ij}\big) \nonumber \\
	& = & \left[\frac{1}{(1+Hr)}\right] \big(\psi_{00} + (1-Hr)n^i\psi_{0i}
	-Hrn^in^j\psi_{ij}\big) \nonumber \\
	\bar{h}_{uA} & = & \left[\frac{e^{-2Hu}}{(1+Hr)^2}\right] \big(r E_A^ih_{0i} -
	Hr^2n^iE^j_Ah_{ij}\big)\nonumber \\
	& = & \big(r E_A^i\psi_{0i} - Hr^2n^iE^j_A\psi_{ij}\big) \nonumber \\
	\bar{h}_{rr} & = & \left[\frac{e^{-2Hu}}{(1+Hr)^4}\right] \big(h_{00} + 2n^ih_{0i} +
	n^in^jh_{ij}\big) \nonumber \\
	& = & \left[\frac{1}{(1+Hr)^2}\right] \big(\psi_{00} + 2n^i\psi_{0i} +
	n^in^j\psi_{ij}\big) \nonumber \\
	\bar{h}_{rA} & = & \left[\frac{e^{-2Hu}\ r}{(1+Hr)^3}\right] \big(E^i_Ah_{0i} +
	n^iE_A^jh_{ij}\big) \nonumber \\
	& = & \left[\frac{r}{(1+Hr)}\right] \big(E^i_A\psi_{0i} + n^iE_A^j\psi_{ij}\big)
	\nonumber \\
	\bar{h}_{AB} & = & \left[\frac{e^{-2Hu}\ r^2}{(1+Hr)^2}\right]
	\big(E_A^iE_B^jh_{ij}\big) 
	~ = ~ r^2 \big(E_A^iE_B^j\psi_{ij}\big) \nonumber 
\end{eqnarray}
Notice that the $\bar{h}_{rr}, \bar{h}_{rA}, \bar{h}_{AB}$ which are needed in determining
the gauge transformation to Bondi gauge, do not have relative $r$ dependence among the
$\psi_{\mu\nu}$ components.

{\em From BS chart $\bar{h}_{\mu\nu}$ to Bondi Gauge $H_{\mu\nu}$:} 

$H_{\mu\nu} = \bar{h}_{\mu\nu} + \hat{\nabla}_{\mu}\xi_{\nu} + \hat{\nabla}_{\nu}\xi_{\mu}$.
\begin{eqnarray}
	H_{uu} & = & \bar{h}_{uu} + 2\partial_{u}\xi_{u} - 2 \hat{\Gamma}_{uu}^{\alpha}
	\xi_{\alpha}
	~ = ~ \bar{h}_{uu} + 2\big(\partial_u\xi_u - H^2r\xi_{u} + H^2r(1-H^2r^2)\xi_r\big)
	; \\
	H_{ur} & = & \bar{h}_{ur} + \partial_u\xi_r + \partial_r\xi_u -
	2\hat{\Gamma}_{ur}^{\alpha}\xi_{\alpha}
	~ = ~ \bar{h}_{ur} + \partial_u\xi_r + \partial_r\xi_u + 2H^2r \xi_r \\
	H_{uA} & = & \bar{h}_{uA} + \partial_u\xi_A + \partial_A\xi_u -
	2\hat{\Gamma}_{uA}^{\alpha}\xi_{\alpha}
	~ = ~ \bar{h}_{uA} + \partial_u\xi_A + \partial_A\xi_u \\
	H_{rr} & = & \bar{h}_{rr} + 2\partial_r\xi_r -
	2\hat{\Gamma}_{rr}^{\alpha}\xi_{\alpha}
	~ = ~ \bar{h}_{rr} + 2\partial_r\xi_r \\
	H_{rA} & = & \bar{h}_{rA} + \partial_r\xi_A + \partial_A\xi_r -
	2\hat{\Gamma}_{rA}^{\alpha}\xi_{\alpha}
	~ = ~ \bar{h}_{rA} + \partial_r\xi_A + \partial_A\xi_r - \frac{2}{r}\xi_A\\
	H_{AB} & = & \bar{h}_{AB} + \partial_A\xi_B + \partial_B\xi_A -
	2\hat{\Gamma}_{AB}^{\alpha}\xi_{\alpha}
	~ = ~ \bar{h}_{AB} +\left(\hat{\nabla}_A \xi_B + \hat{\nabla}_B \xi_A\right)+
	2\hat{\Gamma}_{AB}^u\xi^r
\end{eqnarray}
Noting that $\hat{\Gamma}^u_{AB} = r\gamma_{AB}$, we now replace $\hat{\nabla}_A$ acting on
$\xi_B$ as $D_{A}\xi_B$ where $D_A$ is the covariant derivative corresponding to
$\gamma_{AB}$.

The Bondi gauge conditions require $H_{rr} = 0 = H_{rA} = \hat{g}^{AB}H_{AB}$ i.e.,
\begin{eqnarray}
	\partial_r\xi_r & = & -\frac{1}{2}\bar{h}_{rr} ~ ~ , ~ ~ \partial_A\xi_r =
	-\left(\partial_r - \frac{2}{r}\right)\xi_A -\bar{h}_{rA} ~ \hspace{1.0cm}
	\mbox{and} \\ \label{Xi-rEqn}
	\xi^r & = & -\frac{1}{4r}\gamma^{AB}\big(\bar{h}_{AB} + 2{D}_{A}\xi_{B}\big).  \\
	\label{Xi-RSoln}
	\Rightarrow ~ \xi_u & = & -(1-H^2r^2)\xi^u - \xi^r ~ = ~ (1-H^2r^2)\xi_r - \xi^r
	\nonumber \\
	& = & (1-H^2r^2)\xi_r + \frac{1}{4r}\gamma^{AB}\big(\bar{h}_{AB} +2{D}_A\xi_B \big)
	.  \label{Xi-uEqn}
\end{eqnarray}
Thus, we have first order, inhomogeneous, partial differential equations for $\xi_r$, given
$\xi_A$ while the $\xi_u$ is determined algebraically in terms of $\xi_r, \xi_A$.  The
$\xi_A$ are constrained by the integrability conditions on $\xi_r$.
\begin{eqnarray} \label{IntglCondns1}
	\big[\partial_r, \partial_A\big]\xi_r = 0 & \Rightarrow & \left[\partial^2_r
	-\frac{2}{r}\partial_r + \frac{2}{r^2}\right]\xi_A = -\partial_r\bar{h}_{rA} +
	\frac{1}{2}\partial_A \bar{h}_{rr}  ; \\
	\big[\partial_A, \partial_B\big]\xi_r = 0 & \Rightarrow & \left[\partial_r
	-\frac{2}{r}\right]\zeta_{AB} = -\big(\partial_A\bar{h}_{rB} -
	\partial_B\bar{h}_{rA}\big) ~ , ~ \zeta_{AB} := \big(\partial_A\xi_{B} -
	\partial_B\xi_A\big) .
\end{eqnarray}
Operating by $\partial_B$ on the first integrability condition and antisymmetrizing in $A,B$
we get $\partial_r$ on the second integrability condition, showing their mutual consistency.

To determine the gauge generator $\xi_{\mu}$, we need only the $\bar{h}_{rr}, \bar{h}_{rA}$
and trace of $\bar{h}_{AB}$. Since there are explicit factors depending on $r$, we write
these components as a power series in the following form.  Let,
\begin{equation}
	\psi_{00} := \sum_{k\ge 0} \frac{\psi^{(k)}_{00}(u,\theta,\phi)}{r^k} ~ ~,~ ~
	\psi_{0i} := \sum_{k\ge 0} \frac{\psi_{0i}^{(k)}(u,\theta,\phi)}{r^k} ~ ~,~ ~
	\psi_{ij} := \sum_{k\ge 0} \frac{\psi_{ij}^{(k)}(u,\theta,\phi)}{r^k} \ ;
\end{equation}
The corresponding angular components are defined as: 
\begin{eqnarray}
	\psi_{0r}^{(k)} & := & \psi_{0i}^{(k)}n^i ~ ~,~ ~ \psi_{0A}^{(k)} :=
	\psi_{0i}^{(k)}E^i_A ~ ; \\
	{\psi}^{(k)}_{rr} & := & {\psi}^{(k)}_{ij}n^in^j ~ ~,~ ~ {\psi}^{(k)}_{rA} :=
	{\psi}^{(k)}_{ij}n^iE^j_A ~ ~,~ ~ {\psi}^{(k)}_{AB} := {\psi}^{(k)}_{ij}E^i_AE^j_B
	~.
\end{eqnarray}
\begin{eqnarray}
	\bar{h}_{rr} & = & \frac{1}{(1+Hr)^2}\big(\psi_{00} + 2n^i\psi_{0i} +
	n^in^j\psi_{ij}\big) \nonumber \\
	& := &  \frac{1}{(1+Hr)^2} \sum_{k\ge 0}\frac{\big({\psi}^{(k)}_{rr} + 2
	\psi_{0r}^{(k)} + \psi_{00}^{(k)}\big)}{r^k} ~ \\
	& := & \sum_{k\ge 2}\frac{\bar{h}_{rr}^{(k)}}{r^k} ~ ~ ~ ~ \mbox{with} ~ ~ ~ ~
	\bar{h}_{rr}^{(k)} := \sum_{l=2}^k (-H^{-1})^l(l-1)\big(\psi^{(k-l)}_{rr} +
	2\psi_{0r}^{(k-l)} + \psi_{00}^{(k-l)}\big) \ ; ~ ~ \label{Q1-Coeffs}\\
	\bar{h}_{rA} & = & \frac{r}{1+Hr}\big(E^i_A\psi_{0i} + n^iE^j_A\psi_{ij}\big)
	\nonumber  \\
	& := & \frac{r}{1+Hr} \sum_{k\ge 0}\frac{\big(\psi_{0A}^{(k)} +
	\psi^{(k)}_{rA}\big)}{r^k} ~ , \\
	& := & \sum_{k\ge 0}\frac{\bar{h}^{(k)}_{rA}}{r^k} ~ ~ ~ ~ \mbox{with} ~ ~ ~ ~
	\bar{h}_{rA}^{(k)} := H^{-1}\sum_{l=0}^k (-H^{-1})^l \big(\psi_{0A}^{(k-l)} +
	\psi_{rA}^{(k-l)} \big) \ ; ~ \label{Q2-Coeffs}\\ 
	\bar{h}_{AB} & = & r^2\big(E^i_AE^j_B\psi_{ij}\big) \nonumber \\
	& := & r^2\sum_{k\ge 0}\frac{\psi_{AB}^{(k)}}{r^k} ~ = ~ \sum_{k\ge
	-2}\frac{\psi^{(k+2)}_{AB}}{r^k} ~ \nonumber \\
	& := & \sum_{k\ge -2}\frac{\bar{h}^{(k)}_{AB}}{r^k} ~ ~ ~ ~ \mbox{with} ~ ~ ~ ~
	\bar{h}^{(k)}_{AB} :=\psi^{(k+2)}_{AB},\quad k\ge-2.  \label{Q3-Coeffs}
\end{eqnarray}

We note that to specialize to the synchronous gauge, we just put,
\begin{equation}
	\psi_{00}^{(k)} = 0 = \psi_{0i}^{(k)} ~,~ \psi_{ij}^{(k)} =
	\big[\chi_{ij}^{(k)}\big]^{TT}.  ~ ~ \mbox{for}~ ~ k \ge 0. 
\end{equation}
The solution for the gauge generator and the transformed $H_{AB}$ are entirely given in
terms of the $\bar{h}^{(k)}_{rr}, \bar{h}^{(k)}_{rA}, \bar{h}^{(k)}_{AB}$.

\section{The gauge generator} \label{GaugeGenerator}
The equations determining the gauge generator (\ref{BSDifferentialCondEqn}) are recalled for
convenience of reading,
\begin{eqnarray} 
	\partial_r\xi_r & = & -\frac{1}{2}\bar{h}_{rr} ~ ~ , \nonumber \\
	\partial_r\xi_A - \frac{2}{r}\xi_A & = & -\partial_A\xi_r -\bar{h}_{rA} ~ ~ , \\ 
	\xi^r & = & -\frac{1}{4r}\gamma^{AB}\bar{h}_{AB} - \frac{r}{2}{D}_A\xi^A
\end{eqnarray}
The first two are inhomogeneous, first order differential equations in $(r, x^A)$ for $\xi_r
(= -\xi^{u})$ and $\xi_A (= r^2\xi^A)$ while the last one gives $\xi^r$ in terms of the
integrals of the first two equations.  These in turn determine,
\begin{equation}
	\xi_u ~ = ~ -(1-H^2r^2)\xi^u - \xi^r ~ = ~  (1-H^2r^2)\xi_r +
	\frac{1}{4r}\gamma^{AB}\big(\bar{h}_{AB} + 2D_A\xi_B \big)
\end{equation}
We can view the differential equations as first order equations for $\xi_r$, {\em given}
$\xi_A$ and the $\bar{h}_{rr}, \bar{h}_{rA}$. The integrability conditions for these
equations, give inhomogeneous differential equations for $\xi_A$.  Namely,
\begin{eqnarray} \label{IntglCondns}
	\big[\partial_r, \partial_A\big]\xi_r = 0 & \Rightarrow & \left[\partial^2_r
	-\frac{2}{r}\partial_r + \frac{2}{r^2}\right]\xi_A = -\partial_r\bar{h}_{rA} +
	\frac{1}{2}\partial_A \bar{h}_{rr}  ; \\
	\big[\partial_A, \partial_B\big]\xi_r = 0 & \Rightarrow & \left[\partial_r
	-\frac{2}{r}\right]\zeta_{AB} = -\big(\partial_A\bar{h}_{rB} -
	\partial_B\bar{h}_{rA}\big) ~ , ~ \zeta_{AB} := \big(\partial_A\xi_{B} -
	\partial_B\xi_A\big) .
\end{eqnarray}
Operating by $\partial_B$ on the first integrability condition and antisymmetrizing in $A,B$
we get $\partial_r$ on the second integrability condition, showing their mutual consistency.

In the equations determining the gauge generator, we need only $\bar{h}_{rr}, \bar{h}_{rA},
\bar{h}_{AB}$ which have been expressed in a power series form (\ref{Q1-Coeffs},
\ref{Q2-Coeffs}, \ref{Q3-Coeffs}).  Substitution give,
\begin{eqnarray}
	-\partial_r\bar{h}_{rA} + \frac{1}{2}\partial_A\bar{h}_{rr} & = & \sum_{k\ge
	2}\frac{1}{r^k} \left\{ (k-1)\bar{h}^{(k-1)}_{rA} + \frac{1}{2} \partial_A
	\bar{h}_{rr}^{(k)} \right\} ~ ~ ;  \\
	\partial_A\bar{h}_{rB} - \partial_B\bar{h}_{rA}  & = & \sum_{k\ge 0}\frac{1}{r^k}
	\left\{ \partial_A \bar{h}^{(k)}_{rB} - \partial_B \bar{h}^{(k)}_{rA}\right\} ~ .
\end{eqnarray}

It follows that a similar power series ansatz for the gauge generator can be made:
$\xi_A(u,r,\theta,\phi) =  \sum_{k\ge -2} \Case{\xi^{(k)}_{A} (u,\theta,\phi)} {r^k}$.  The
first of the integrability conditions (\ref{IntglCondns}) leads to,
\begin{equation}
	\sum_{k\ge -2}\frac{(k+1)(k+2)}{r^{k+2}}\xi^{(k)}_A  =  
	\sum_{k\ge 0}\frac{(k)(k-1)}{r^{k}}\xi^{(k-2)}_A ~ = ~ 
	\sum_{k\ge 2}\frac{1}{r^k} \left\{ (k-1)\bar{h}^{(k-1)}_{rA} + \frac{1}{2}
	\partial_A \bar{h}_{rr}^{(k)} \right\} ~ ; 
\end{equation}
The coefficients $\xi_A^{(-2)}, \xi_A^{(-1)}$ are undetermined.  They correspond to the
homogeneous solution of the integrability condition and are part of the residual coordinate
freedom of the Bondi gauge. The coefficients $\xi_A^{(k\ge 0)}$ are determined as,
\begin{eqnarray}
	\xi_A^{(k)} & = & \frac{1}{(k+1)(k+2)}\left((k+1)\bar{h}^{(k+1)}_{rA} +
	\frac{1}{2}\partial_A\bar{h}_{rr}^{(k+2)}\right) ~ \forall ~ k \ge 0 \ .
\end{eqnarray}
The second integrability condition leads to,
\begin{equation}
	\sum_{k\ge 0}\frac{k+1}{r^k} \big(\partial_A\xi^{(k-1)}_{B} -
	\partial_B\xi^{(k-1)}_A\big) 
	~ = ~ \sum_{k\ge 0}\frac{\partial_A \bar{h}^{(k)}_{rB} - \partial_B
	\bar{h}^{(k)}_{rA}}{r^k} ~ . 
\end{equation}
This holds identically for $k \ge 1$ i.e. for $\xi_A^{(k\ge 0)}$ while $\xi_A^{(-1)}$
satisfies,
\begin{equation}
	\partial_A\xi_B^{(-1)} - \partial_B\xi_A^{(-1)} ~ = ~ \big(
	\partial_A\bar{h}^{(0)}_{rB} - \partial_B\bar{h}^{(0)}_{rA} \big).
\end{equation}
This completes the solution for $\xi_{A}$.

Going back to the $\xi_r$ equation, we see that it too has a power series form. Thus let
$\xi_r(u,r,\theta,\phi) := \sum_{k\ge 0}\Case{\xi_r^{(k)}(u,\theta,\phi)}{r^k}.$
Substitution in the radial equation (\ref{Xi-rEqn}) leads to,
\begin{eqnarray}
	\xi_r^{(k)} & = & \frac{1}{2k} \bar{h}_{rr}^{(k+1)}  ~ ~ \forall ~ ~ k \ge 1 ~ ~
	\mbox{and} ~  ~ \xi^{(0)}_r ~ \mbox{is free.}
\end{eqnarray}
Substitution in the angular equation for $\xi_r$ leads to a condition on the free
coefficient $\xi_r^{(0)}$
\begin{eqnarray}
	\partial_A\sum_{k\ge 0} \frac{1}{r^k} \xi_r^{(k)} & = & \partial_A \xi_r^{(0)} +
	\sum_{k\ge 1} \frac{1}{r^k} \Big( \frac{\partial_A \bar{h}^{(k+1)}_{rr}}{2k}\Big) \\
	& = & - \sum_{k\ge 0} \frac{\bar{h}_{rA}^{(k)}}{r^k} + \sum_{k\ge
	0}\frac{(k+1)\xi_A^{(k-1)}}{r^k}  ~ ~ \Rightarrow ~ \\
	\partial_A \xi_r^{(0)} & = & - \bar{h}_{rA}^{(0)} + \xi_A^{(-1)} \ .
\end{eqnarray}
The $k \ge 1$ equations are satisfied identically. Thus $\xi_r$ is also determined with the
coefficient $\xi^{(0)}_r$ partially restricted.

The component $\xi_u$ is obtained algebraically and is given by,
\begin{equation}
	\xi_u ~ = ~ (1- H^2r^2)\xi_r + \frac{1}{4r} \gamma^{AB} \big(\bar{h}_{AB} +
	2{D}_A\xi_B \big)  ~ ~ := ~ ~ \sum_{k\ge -2}\frac{\xi_u^{(k)}(u,\theta,\phi)}{r^k} \
	.
\end{equation}
The covariant derivatives act only on $\xi_A^{(k)}$ and we get,
\begin{eqnarray}
	\xi_{u}^{(-2)} & = & -H^2\xi_r^{(0)} ; \\
	\xi_{u}^{(-1)} & = & -H^2\xi_r^{(1)} + \frac{1}{4}\gamma^{AB}\big(
	\bar{h}_{AB}^{(-2)} + 2 {D}_A\xi_B^{(-2)} \big) ; \\
	\xi_u^{(k\ge 0)} & = & \xi_r^{(k)} - H^2\xi_r^{(k+2)} + \frac{1}{4}\gamma^{AB}\big(
	\bar{h}_{AB}^{(k-1)} + 2 {D}_A\xi_B^{(k-1)} \big) . 
\end{eqnarray}

We have the gauge generator. The calculation of the remaining components of $H_{\mu\nu}$ is
now straightforward. All are again power series. For instance, using $\hat{\Gamma}^u_{AB} =
r\gamma_{AB}$ and $\xi^r = -\Case{1}{4r} \gamma^{CD} (\bar{h}_{CD}+2{D}_C\xi_D)$ we get 
\begin{eqnarray}
	H_{AB} & = & \bar{h}_{AB} + {D}_A\xi_B + {D}_B\xi_A + 2\hat{\Gamma}^u_{AB}\xi^r ~ \\
	& = & \bar{h}_{AB} + {D}_A\xi_B + {D}_B\xi_A
	-\frac{1}{2}\gamma_{AB}{\gamma}^{CD}\big( \bar{h}_{CD} +2 {D}_C\xi_D \big) ~ ~ := ~
	~ \sum_{k\ge -2} \frac{H_{AB}^{(k)}}{r^k} ~ \Rightarrow \\
	H_{AB}^{(k)} & = & \bar{h}_{AB}^{(k)} + \big( {D}_A\xi_B^{(k)} + {D}_B \xi_A^{(k)}
	\big) -\frac{1}{2} \gamma_{AB}\gamma^{CD} \big( \bar{h}_{CD}^{(k)} + 2
	{D}_C\xi_D^{(k)} \big) ~ \forall ~ k \ge -2 \ .
\end{eqnarray}
Other components have similar power series form.
\section{Calculation of $D_{AB}$} \label{CalnD-AB}
Introduce an auxiliary notation,
\begin{eqnarray} \label{MuABDefn}
	T^{(k)}_{AB} := \bar{h}^{(k)}_{AB} + {D}_A\xi^{(k)}_B + {D}_B\xi^{(k)}_A ~ ~
	\mbox{so that} ~ ~ H_{AB}^{(k)} ~ = ~ T_{AB}^{(k)} - \frac{1}{2}\gamma_{AB}T_C^{~C}
	\ . 
\end{eqnarray}
The $k = -2, -1$ coefficients contain the free $\xi_{A}^{(-2)}, \xi_A^{(-1)}$ coefficients.
The coefficient for $k = -1$ gives $C_{AB} = H_{AB}^{(-1)}$ while the $D_{AB} =
H_{AB}^{(0)}$.

For $k\ge 0$ the $\xi^{(k)}_A$ are determined in terms of the $\bar{h}'s$ and so are the
$H_{AB}$'s. Substituting for $\xi_A^{(k)}$, we write
\begin{eqnarray}
	\xi_A^{(k)} & = & \frac{1}{(k+2)} \bar{h}_{rA}^{(k+1)} +
	\frac{1}{2(k+1)(k+2)}\partial_A \bar{h}_{rr}^{(k+2)} ~ ~ \Rightarrow \\
	T^{(k)}_{AB} & = & \bar{h}_{AB}^{(k)} + \frac{1}{(k+2)} \big({D}_A
	\bar{h}_{rB}^{(k+1)} + {D}_B \bar{h}_{rA}^{(k+1)}\big) \nonumber \\
	& & \hspace{1.2cm} + \ \frac{1}{2(k+1)(k+2)} \big({D}_A {D}_B + {D}_B {D}_A\big)
	\bar{h}_{rr}^{(k+2)} \ . 
\end{eqnarray}
For notational convenience, we introduce {\em scaled versions} of $\bar{h}^{(k)}_{**}$,
namely,
\begin{eqnarray}
	& & \tau^{(k)}_{AB} := \bar{h}^{(k)}_{AB} ~ , ~ \tau^{(k)}_{rA} := \frac{1}{k+2}
	\bar{h}^{(k+1)}_{rA} ~ , ~ \tau^{(k)}_{rr} := \frac{1}{(k+1)(k+2)}
	\bar{h}^{(k+2)}_{rr} \ ; \nonumber \\
	& & \hspace{1.2cm} \leftrightarrow T_{AB}^{(k)} = \tau^{(k)}_{AB} + {D}_{A}
	\tau^{(k)}_{rB} + {D}_{B}\tau^{(k)}_{rA} + {D}_{(A} {D}_{B)} \tau^{(k)}_{rr}
	\label{TauDefns}
\end{eqnarray}
At this stage, we need to express the $\bar{h}^{(k)}$'s back in terms of the Cartesian
components.  Each term in the $\bar{h}^{(k)}$'s has the form
$[\tilde{S}_{**|L}^{(k)}(u)\cdot n^{i_1}\cdots n^{i_L}(\theta,\phi)]$ where $(**)$ denotes
$(00), (0i), (ij)$.  The $\bar{h}^{(k)}$'s are thus  polynomials in the unit radial vector
whose coefficients are functions of $u$ alone.  The entire expression for each of the
$\tau^{(k)}$'s is a polynomial in basis vectors $n^i, E^i_A$.  We need two derivatives $D_A$
on $n^in^j$ from $\bar{h}_{rr}^{(k)}$ and one derivative on $n^iE^j_A$ from $\bar{h}_{rA}$.
We also need the angular derivatives of the $\bar{h}^{(k)}_{ij}$.  The action of the
derivatives is simple on each of these: $D_A n^i = E^i_A, D_A E^j_B = -\gamma_{AB} n^j$.

To revert to the Cartesian components, read off the general expressions for the $\tau$'s
from the definitions of $\bar{h}_{AB}$ (\ref{Q3-Coeffs}), $\bar{h}_{rA}$ (\ref{Q2-Coeffs})
and $\bar{h}_{rr}$ (\ref{Q1-Coeffs}). Note that $\psi_{ij}, \psi_{0j}$ occur in both
$\bar{h}_{rr}, \bar{h}_{rA}$ and have different $\sum_{l}$. So an additional labels are
introduced below. We have,
\begin{eqnarray}
	\tau^{(k)}_{AB} & = & E^i_AE^j_B\tau^{(k)}_{3ij} ~,~ \tau^{(k)}_{3ij} :=
	\psi^{(k+2)}_{ij} ;  \\
	\tau^{(k)}_{rA} & = & n^iE^j_A\tau^{(k)}_{2ij} + E^j_A\tau^{(k)}_{2j} ~; \\
	& & \tau^{(k)}_{2ij} := \frac{H^{-1}}{k+2}\sum_{l=0}^{k+1}
	\big(-H^{-1}\big)^l\psi^{(k+1-l)}_{ij} ~ , \nonumber \\
	& & \tau^{(k)}_{2j} := \frac{H^{-1}}{k+2}\sum_{l=0}^{k+1}
	\big(-H^{-1}\big)^l\psi^{(k+1-l)}_{0j} \nonumber \\
	\tau^{(k)}_{rr} & = & n^in^j\tau^{(k)}_{1ij} + 2n^i\tau^{(k)}_{1i} + \tau^{(k)}_{1}
	~ ; \\
	& & \tau^{(k)}_{1ij} := \frac{1}{(k+1)(k+2)}\sum_{l=2}^{k+2} \big(-H^{-1}\big)^l
	(l-1) \psi^{(k+2-l)}_{ij} ~ , \nonumber \\
	& & \tau^{(k)}_{1i} := \frac{1}{(k+1)(k+2)}\sum_{l=2}^{k+2} \big(-H^{-1}\big)^l
	(l-1) \psi^{(k+2-l)}_{0i} \nonumber \\
	& & \tau^{(k)}_{1} := \frac{1}{(k+1)(k+2)}\sum_{l=2}^{k+2} \big(-H^{-1}\big)^l (l-1)
	\psi^{(k+2-l)}_{00} \nonumber 
\end{eqnarray}

It follows,
\begin{eqnarray}
	D_A\tau^{(k)}_{rB} & = & D_A(\tau^{(k)}_{2ij}n^iE^j_B + E^j_B\tau_{2j})  , \nonumber
	\\
	& = & \Big\{E^i_AE^j_B\tau^{(k)}_{2ij} + n^i(-\gamma_{AB}n^j)\tau^{(k)}_{2ij} +
	n^iE^j_B D_A \tau^{(k)}_{2ij}\Big\} \nonumber \\
& & \hspace{2.1cm} + ~ \Big\{-\gamma_{AB}n^j\tau^{(k)}_{2j} + E^j_B D_{A} \tau^{(k)}_{2j}
\Big\}\ ;
\end{eqnarray}
Hence,
\begin{eqnarray}
	D_A\tau^{(k)}_{rB} + D_B\tau^{(k)}_{rA} & = & E^i_AE^j_B\left\{2\tau^{(k)}_{2ij}
	\right\} - 2\gamma_{AB} \Big\{n^in^j \tau^{(k)}_{2ij} + n^i \tau^{(k)}_{2i} \Big\}
	\nonumber \\
	& & + n^i\big( E^j_BD_A + E^j_AD_B\big) \tau^{(k)}_{2ij} + \big( E^j_BD_A +
	E^j_AD_B\big) \tau^{(k)}_{2j} .
\end{eqnarray}

Lastly, for $\tau^{(k)}_{rr}$,
\begin{eqnarray}
	{D}_B\big(n^in^j\tau^{(k)}_{1ij}\big) & = & \big(E^i_Bn^j +
	E^j_Bn^i\big)\tau^{(k)}_{1ij} + n^in^jD_B \tau^{(k)}_{1ij} \ ; \\
	{D}_A{D}_B \big(n^in^j\tau^{(k)}_{1ij}\big) & = & 
	\big(2E^i_AE^j_B -2\gamma_{AB}n^in^j\big)\tau^{(k)}_{1ij} + n^in^j D_AD_B
	\tau^{(k)}_{1ij} \nonumber \\
	& & + \Big\{n^j\big(E^i_AD_B + E^i_BD_A\big) + n^i\big(E^j_AD_B +
	E^j_BD_A\big)\Big\} \tau^{(k)}_{1ij} \nonumber \\ 
	\therefore {D}_{(A}{D}_{B)}\big(n^in^j\tau^{(k)}_{1ij}\big) & = & 2\big(E^i_AE^j_B
	-\gamma_{AB}n^in^j\big)\tau^{(k)}_{1ij} + n^in^j D_{(A}D_{B)} \tau^{(k)}_{1ij}
	\nonumber \\
	& & + \Big\{n^j\big(E^i_AD_B + E^i_BD_A\big) + n^i\big(E^j_AD_B +
	E^j_BD_A\big)\Big\} \tau^{(k)}_{1ij} \nonumber \\ 
	{D}_B\big(2n^i\tau^{(k)}_{1i} + \tau^{(k)}_{1}\big) & = & 2E_B^i\tau^{(k)}_{1i} +
	2n^iD_B \tau^{(k)}_{1i} + D_B \tau^{(k)}_{1} \ . \\
	{D}_A{D}_B\big(2n^i\tau^{(k)}_{1i} + \tau^{(k)}_{1} \big) & = & 2\big(E^i_BD_A +
	E^i_AD_B\big)\tau^{(k)}_{1i} -2\big(\gamma_{AB}n^i\tau^{(k)}_{1i}\big) \nonumber \\
	& & + 2n^i D_AD_B \tau^{(k)}_{1i} + D_AD_B \tau^{(k)}_{1} \\
	\therefore {D}_{(A}{D}_{B)}\big(2n^i\tau^{(k)}_{1i} + \tau^{(k)}_{1}\big) & = &
	2\big(E^i_BD_A + E^i_AD_B\big)\tau^{(k)}_{1i}
	-2\big(\gamma_{AB}n^i\tau^{(k)}_{1i}\big) \nonumber \\
	& & + 2n^i D_{(A}D_{B)} \tau^{(k)}_{1i} + D_{(A}D_{B)} \tau^{(k)}_{1} .
\end{eqnarray}

Putting all together,
\begin{eqnarray}
	T^{(k)}_{AB} & = & \tau^{(k)}_{AB} + \big({D}_A\tau^{(k)}_{rB} +
	{D}_B\tau^{(k)}_{rA}\big) + {D}_{(A}{D}_{B)}\tau^{(k)}_{rr} \nonumber \\ 
	& = & \Big[ E^i_AE^j_B \Big\{\tau^{(k)}_{3ij} + 2\tau^{(k)}_{2ij} + 2
	\tau^{(k)}_{1ij} \Big\} \Big]_{\textcircled{\scriptsize{1}}} \nonumber \\
& & + \Big[ n^i\big(E^j_AD_B + E^j_BD_A\big)\tau^{(k)}_{2ij}
\Big]_{\textcircled{\scriptsize{2}}} \nonumber \\
& & + \Big[ \Big\{n^i\big(E^j_AD_B + E^j_BD_A\big) + n^j\big(E^i_AD_B + E^i_BD_A\big) \Big\}
\tau^{(k)}_{1ij}\Big]_{\textcircled{\scriptsize{3}}} \nonumber \\
& &  + \Big[ \big(E^j_AD_B + E^j_BD_A\big)\tau^{(k)}_{2j} + 2\big( E^j_AD_B + E^j_BD_A\big)
\tau^{(k)}_{1j} \Big]_{\textcircled{\scriptsize{4}}} \nonumber \\
& &  + \Big[ n^in^jD_{(A}D_{B)} \tau^{(k)}_{1ij} + 2n^iD_{(A}D_{B)} \tau^{(k)}_{1i} +
D_{(A}D_{B)} \tau^{(k)}_{1} \Big]_{\textcircled{\scriptsize{5}}} \nonumber\\
& &  -\gamma_{AB}\Big[ 2n^in^j\big(\tau^{(k)}_{2ij} + \tau^{(k)}_{1ij}\big) + 2n^i
\big(\tau^{(k)}_{2i} + \tau^{(k)}_{1i}\big) \Big] \ .  ~ \label{T-AB-General}
\end{eqnarray}

{\em Note:} The above expressions are general expressions transforming $\psi_{\mu\nu}$
expressed as a power series into the Bondi-Sachs gauge.  Whether they are linearized
solutions or not has not been used anywhere.  We can choose them to be the solutions in the
generalized harmonic gauge. The above expressions can be adopted to the synchronous gauge by
dropping the $\tau_{*i}, \tau_{*}$ which come from the $\psi_{00}, \psi_{0i}$ and using
$\psi_{ij} = [\psi_{ij}]^{TT}$. In this case the $\tau_{*ij}$ are traceless.

Now specialize to the $D_{AB} = H^{(0)}_{AB}$. In the following, we will drop the
superscript $(k)$. The vanishing of $D_{AB}$ follows if $T_{AB} \propto \gamma_{AB}$.  This
means that the sum of the first 5 square brackets must be proportional to $\gamma_{AB}$ 
\begin{equation} \label{MasterCondn}
	\sum_{a=1}^5 \big[\cdots\big]_{AB} ~ \propto ~ \gamma_{AB} \ .
\end{equation}
\subsection{Verification for Comp\`ere et al truncation}
For $D_{AB}$ we have,
\begin{eqnarray}
	\tau_{3ij} = \psi_{ij}^{(2)} ~ , ~ \tau_{2ij} = \frac{1}{2H}\big(\psi^{(1)}_{ij} -
	H^{-1}\psi_{ij}^{(0)}\big) ~ , ~ \tau_{1ij} = \frac{1}{2H^2}\psi_{ij}^{(0)}  \\
	\tau_{1j} = \frac{1}{2H^2}\psi_{0j}^{(0)} ~ , ~ \tau_{2j} =
	\frac{1}{2H}\big(\psi_{0j}^{(1)} - H^{-1}\psi_{0j}^{(0)}\big) ~ , ~ \tau_1 =
	\frac{1}{2H^2}\psi_{00}^{(0)} \ .
\end{eqnarray}
We write the {\em five}  groups of terms in the square bracket in equation
(\ref{T-AB-General}) as:
\begin{eqnarray}
	\Big[E^i_AE^j_B\left\{\tau_{3ij} + 2\tau_{2ij} +
	2\tau_{1ij}\right\}\Big]_{\textcircled{\scriptsize{1}}} & = &
	E^i_AE^j_B\left\{\psi_{ij}^{(2)} + \frac{1}{H} \psi_{ij}^{(1)}\right\} ; \nonumber
	\\
	2\Big[ n^iE^j_{(A}D_{B)}\tau_{2ij} \Big]_{\textcircled{\scriptsize{2}}} & = &
	\frac{1}{H} \Big\{ n^i E^j_{(A} D_{B)} \Big\} \Big(\psi^{(1)}_{ij} -
	H^{-1}\psi^{(0)}_{ij}\Big) ; \nonumber \\
	4\Big[ \Big\{n^{(i}E^{j)}_{(A}D_{B)} \Big\}
	\tau_{1ij}\Big]_{\textcircled{\scriptsize{3}}} & = & \frac{2}{H^2}
	\Big\{n^{(i}E^{j)}_{(A}D_{B)} \Big\} \psi^{(0)}_{ij} ; \nonumber \\
	\Big[ 2E^j_{(A}D_{B)} \tau_{2j} \Big]_{\textcircled{\scriptsize{4}}} & = &
	\frac{1}{H} \Big\{ E^j_{(A}D_{B)}\Big\}\Big(\psi^{(1)}_{0j} -
	H^{-1}\psi^{(0)}_{0j}\Big) ; \nonumber \\
	\Big[ n^in^jD_{(A}D_{B)} \tau_{1ij} \Big]_{\textcircled{\scriptsize{5}}} & = &
	\frac{1}{2H^2} \Big\{ n^in^jD_{(A}D_{B)} \Big\} \psi^{(0)}_{ij} \ .
	\label{5GroupsEqn}\\
	\Big[ 4E^j_{(A}D_{B)} \tau_{1j} \Big] & = &  \frac{2}{H^2} \Big\{
	E^j_{(A}D_{B)}\Big\} \psi^{(0)}_{0j} ; \nonumber \\
	\Big[ 2n^i D_{(A}D_{B)} \tau_{1i} + D_{(A}D_{B)} \tau_{1} \Big] & = & \frac{1}{2H^2}
	\Big\{ 2n^iD_{(A}D_{B)} \psi^{(0)}_{0i} + D_{(A}D_{B)} \psi^{(0)}_{00} \Big\}
	.\nonumber  
\end{eqnarray}
The last two equations are separated from the groups \textcircled{4} and \textcircled{5} as
they will turn out to vanish. The sum of the five groups of terms should have the required
form (\ref{MasterCondn}). 

Recall that $\psi_{\mu\nu}$ are the scaled components of the physical perturbation, whereas
the displayed solutions, including those of \cite{CHK}, are trace reversed. Their relation
is given in (\ref{PhysicalPerturbationReconstruction}).

As per the quadrupolar solution given in \cite{CHK}, $\hat{\chi}, \chi_{0i}$ do not have
$r^0$ term. Hence, $\psi_{00}^{(0)} = 0$.  Hence,
\begin{equation}
	\tau_{1j} = 0 ~ , ~ \tau_{2j} = \frac{\chi_{0j}^{(1)}}{2H} ~ , ~ \tau_1 = 0 \ .
\end{equation}

The requisite terms from the solution are:
\begin{eqnarray}
	\psi_{0i}^{(1)} & = & {\chi}_{0i}^{(1)} \nonumber \\
	& = & 4\Big[ \Big\{P_i - \frac{H}{2} (\partial_t - H)P_{i|ll}\Big\} + \Big\{ -
	n^kS_{ik} \Big\} + \Big\{ \frac{n^kn^l}{2} (\partial_t + H)S_{kl|i}\Big\} \Big] ; \\
\psi_{ij}^{(0)} & = & {\chi}_{ij}^{(0)} - \delta_{ij}{\chi}^{(0)}_{ll} \nonumber \\
& = & 4H^2\Big[ \Big\{ -P_{(i|j)} -\frac{H}{2}S_{ij|ll}\Big\} + \Big\{ n^kS_{ij|k} \Big\} +
\Big\{\frac{n^kn^l \partial_t S_{ij|kl}}{2} \Big\} \Big] \nonumber \\
& & - \delta_{ij} 4H^2\Big[ \Big\{ -P_{(l|l)} -\frac{H}{2}S_{|ll}\Big\} + \Big\{ n^kS_{|k}
\Big\} + \Big\{\frac{n^kn^l \partial_t S_{|kl}}{2} \Big\} \Big] ; \nonumber \\
\psi^{(1)}_{ij} & = & {\chi}^{(1)}_{ij} + \delta_{ij}\Big(\frac{{\hat{\chi}}^{(1)}}{2} -
{\chi}^{(1)}_{ll}\Big) \nonumber \\
& = & 4\Big[ \Big\{ S_{ij} - \frac{H\partial_t S_{ij|ll}}{2} \Big\} + \Big\{ n^k\partial_t
S_{ij|k} \Big\} + \Big\{\frac{n^kn^l \partial^2_t S_{ij|kl}}{2} \Big\} \Big] \\ 
& & + \delta_{ij}\Big[ 2\Big(Q^{(\rho+p)} - \frac{H}{2}(\partial_t -H)Q^{(\rho+p)}_{ll} +
	Q^{(\rho+p)}_{k}n^k + \frac{1}{2}(\partial^2_t - H^2)Q^{(\rho+p)}_{kl}n^kn^l\Big)
	\nonumber \\
	& & ~ ~ ~ ~ -4\Big( \Big\{ S - \frac{H}{2}\partial_t S_{|ll}\Big\} +
\Big\{n^k\partial_tS_{|k}\Big\} + \Big\{\frac{n^kn^l\partial_t^2S_{|kl}}{2}\Big\} \Big)
\Big] ; \nonumber \\ 
\psi^{(2)}_{ij} & = & {\chi}^{(2)}_{ij} + \delta_{ij} \Big( \frac{{\hat{\chi}}^{(2)}}{2} -
{\chi}^{(2)}_{ll}\Big) \nonumber \\
& = & 4\Big[ \Big\{-\frac{\partial_t S_{ij|ll}}{2} \Big\} + \Big\{n^kS_{ij|k}\Big\} + \Big\{
\frac{3n^kn^l \partial_t S_{ij|kl}}{2}\Big\} \Big] \\
& & + \delta_{ij}\Big[ 2\Big(Q^{(\rho+p)}_{|k}n^k + \frac{1}{2}\big(3n^kn^l
	-\delta^{kl}\big) \partial_tQ^{(\rho+p)}_{|kl}\Big) 
	-4\Big(S_{|k}n^k + \frac{\big(3n^kn^l - \delta^{kl}\big)\partial_t S_{|kl}}{2}\Big)
\Big] \ . \nonumber 
\end{eqnarray}

Substitution in the (\ref{5GroupsEqn}) leads to:
\begin{eqnarray}
	\textcircled{\scriptsize{1}} & : & E^i_AE^j_B\Big[ ~ \frac{4S_{ij}}{H} +
		4n^k\big(S_{ij|k} + H^{-1}\dot{S}_{ij|k}\big) \nonumber \\
	& & \hspace{1.3cm} + n^kn^l\big( 6 \dot{S}_{ij|kl} + 2H^{-1}\ddot{S}_{ij|kl}\big) -
4\dot{S}_{ij|ll}\Big]  \\
\textcircled{\scriptsize{2}} & : & 4E^i_{(A}E^j_{B)}\Big[ n^k\big(-S_{kj|i} +
H^{-1}\dot{S}_{kj|i}\big) - n^kn^l\big(\dot{S}_{kj|il} - H^{-1}\ddot{S}_{kj|il}\big)\Big] \\
\text{\textcircled{\scriptsize 4}} & : & E^i_{(A}E^j_{B)}\Big[\frac{4}{H} \left\{-S_{ij} +
n^l(\partial_t + H) S_{l(i|j)}\right\} \Big] \\
\text{\textcircled{\scriptsize 3}} & : & 4E^i_{(A}E^j_{B)}\Big[ n^k\Big(S_{ki|j} +
S_{kj|i}\Big) + n^kn^l\Big( \dot{S}_{ki|jl} + \dot{S}_{kj|il}\Big) \Big] \\
\textcircled{\scriptsize{5}} & : & E^i_{(A}E^j_{B)}\Big[ 2 \big( n^kn^l \dot{S}_{kl|ij} -
\dot{S}_{|ij} \big) \Big] 
\end{eqnarray}
The terms which are proportional to $\gamma_{AB}$ have already been dropped.  Notice that
there are no terms with $P_{i|j}, Q^{(\rho+p)}_{*}$. Through the consistency with the source
conservation, the quadrupolar truncation used also imposes conditions on the stress moments,
namely, $S_{(ij|k)} = 0\ , S_{i(j|kl)} = 0$. These in turn imply $S_{(ij|k)l} = 0$ and
$S_{ij|kl} = S_{kl|ij}$.

Using these properties of the stress moments, it follows that the sum of the five groups of
terms reduces to,
\begin{equation}
	\mbox{sum} ~ = ~ 6E^i_AE^j_B\Big( \dot{S}_{ij|kl}(n^kn^l -\delta^{kl}) \Big) ~ = ~
	-6E^i_AE^j_B E^k_C E^l_D\gamma^{CD} \Big( \dot{S}_{ij|kl}\Big) .
\end{equation}
It looks like the leftover term should vanish implying a further condition on the stress
moments. However, {\em this term is also proportional to $\gamma_{AB}$} and we verify that
$D_{AB} = 0$ holds for the consistent quadrupolar truncation of \cite{CHK}.

A direct argument follows by defining $\Sigma_{ABCD}:=E_A^iE_B^jE_C^kE_D^l\dot S_{ij|kl}$.
Pair symmetry and the cyclic condition imply, in two angular dimensions,
\begin{equation}
	\Sigma_{ABCD} =f\big(\gamma_{AC}\gamma_{BD}+\gamma_{AD}\gamma_{BC}
	-2\gamma_{AB}\gamma_{CD}\big),\qquad \gamma^{CD}\Sigma_{ABCD}=-2f\gamma_{AB}.
\end{equation}
Hence the remainder is pure trace and its trace-free part vanishes, without an additional
condition on the stress moments.
\subsection{Verification for Alternate truncation}
The five groups of terms are still those displayed in (\ref{5GroupsEqn}); only the truncated
solution changes.  The solution is given in (\ref{Chi-ijSoln}) and (\ref{Chi-0iSoln}), with
the mass and current moments related to the stress moments by (\ref{MomentConservEqn}).  The
physical/trace-reversed distinction cannot be discarded completely. Since
\(\hat\chi^{(0)}=0\), the scalar part of the physical spatial perturbation contributes to
group 5 and must be retained (TF below stands for trace free):
\begin{equation}
	\psi_{ij}^{(0)} =\chi_{ij}^{(0)}-\delta_{ij}\chi_{aa}^{(0)},\qquad
	\Delta\big[\text{group 5}\big]_{AB}^{\mathrm{TF}} =-2\big[E_A^iE_B^j\dot
	S_{aa|ij}\big]^{\mathrm{TF}} .
\end{equation}

Writing a dot for $\partial_t$, the coefficients needed from (\ref{Chi-ijSoln}) are
\begin{equation}
	\begin{aligned}
		\chi_{ij}^{(0)}={}& \chi_{ij}(-\infty)
		+4H^2\left[-P_{(i|j)}-\frac{H}{2}S_{ij|aa} +n^kS_{ij|k}+\frac12n^kn^l\dot
		S_{ij|kl}\right],\\
		\chi_{ij}^{(1)}={}& 4\left[S_{ij}-\frac{H}{2}\dot S_{ij|aa} +n^k\dot
		S_{ij|k}+\frac12n^kn^l\ddot S_{ij|kl}\right],\\
	\chi_{ij}^{(2)}={}& 4\left[-\frac12\dot S_{ij|aa} +n^kS_{ij|k}+\frac32n^kn^l\dot
	S_{ij|kl}\right],\\
	\psi_{0i}^{(0)}={}&0,\qquad \hat\chi^{(0)}=0.
\end{aligned}
\label{AlternateSpatialCoefficients}
\end{equation}
The angle-independent term $\chi_{ij}(-\infty)$ drops out when the angular derivatives are
taken.  The scalar term entering the physical spatial perturbation is
$\hat\chi^{(0)}-\chi_{aa}^{(0)}=-\chi_{aa}^{(0)}$ and is responsible for the final $-\dot
S_{aa|ij}$ term in group 5.

Using the $L=1,2,3$ cases of the conservation equations (\ref{MomentConservEqn}) repeatedly
in (\ref{Chi-0iSoln}), the coefficient needed from the vector component becomes
\begin{equation}
	\begin{aligned}
		\psi_{0i}^{(1)}={}&4P_i+4H S_{ia|a}-4n^aS_{ia}\\
		&-2n^an^b\left[H\left(S_{ia|b}+S_{ib|a}\right)
		+\partial_t\left(S_{ia|b}+S_{ib|a}\right)\right]\\
		&+2Hn^a\partial_t\left(S_{ia|bb}+2S_{ib|ab}\right)\\
		&-\frac{4H}{3}n^an^bn^c\partial_t
		\left(S_{ia|bc}+S_{ib|ac}+S_{ic|ab}\right)\\
		&-\frac{2}{3}n^an^bn^c\partial_t^2
		\left(S_{ia|bc}+S_{ib|ac}+S_{ic|ab}\right).
	\end{aligned}
	\label{AlternateVectorCoefficient}
\end{equation}
The terms $4P_i$ and $4H S_{ia|a}$ have no angular dependence.  Since group
\textcircled{\scriptsize 4} contains an angular derivative of $\psi_{0i}^{(1)}$, they obey
$D_A P_i=0=D_A S_{ia|a}$ and do not contribute to that group.

Substitution in the five groups of (\ref{5GroupsEqn}), with terms proportional to
$\gamma_{AB}$ suppressed, gives
\begin{equation}
	\begin{aligned}
		\text{\textcircled{\scriptsize 1}}:\quad
		&E_A^iE_B^j\Bigg[\frac{4}{H}S_{ij}-4\dot S_{ij|aa}
			+4n^k\left(S_{ij|k}+\frac{1}{H}\dot S_{ij|k}\right)\\
		&\hspace{2.6cm} +n^kn^l\left(6\dot S_{ij|kl}+\frac{2}{H}\ddot
	S_{ij|kl}\right) \Bigg],\\[1mm] \text{\textcircled{\scriptsize 2}}:\quad
	&E_A^iE_B^j\Bigg[-2n^k\left(S_{kj|i}+S_{ki|j}\right)
		+\frac{2}{H}n^k\partial_t\left(S_{kj|i}+S_{ki|j}\right)\\
	&\hspace{2.6cm} -2n^kn^l\partial_t\left(S_{kj|il}+S_{ki|jl}\right)
+\frac{2}{H}n^kn^l\partial_t^2 \left(S_{kj|il}+S_{ki|jl}\right)\Bigg],\\[1mm]
\text{\textcircled{\scriptsize 3}}:\quad &4E_A^iE_B^j\Bigg[n^k\left(S_{kj|i}+S_{ki|j}\right)
+n^kn^l\partial_t\left(S_{kj|il}+S_{ki|jl}\right)\Bigg],\\[1mm]
\text{\textcircled{\scriptsize 4}}:\quad &E_A^iE_B^j\Bigg[-\frac{4}{H}S_{ij}
	+2\partial_t\left(S_{ij|aa}+S_{ja|ia}+S_{ia|ja}\right)\\
	&\hspace{1.0cm} -\frac{2}{H}n^k\Big\{H\left(2S_{ij|k}+S_{jk|i}+S_{ik|j}\right)
	+\partial_t\left(2S_{ij|k}+S_{jk|i}+S_{ik|j}\right)\Big\}\\
	&\hspace{1.0cm} -4n^kn^l\partial_t \left(S_{ij|kl}+S_{jk|il}+S_{ik|jl}\right)\\
&\hspace{1.0cm} -\frac{2}{H}n^kn^l\partial_t^2
\left(S_{ij|kl}+S_{jk|il}+S_{ik|jl}\right)\Bigg],\\
\hspace{1mm} \text{\textcircled{\scriptsize 5}}:\quad &2E_A^iE_B^j\left[n^kn^l\dot
S_{kl|ij}-\dot S_{aa|ij}\right].
\end{aligned}
\label{AlternateFiveGroups}
\end{equation}
All terms containing $S_{ij}$, $S_{ij|k}$, $\dot S_{ij|k}$ and $\ddot S_{ij|kl}$ cancel when
these five expressions are added.  Their tangential trace-free part is therefore
\begin{equation}
	\begin{aligned}
		D_{AB} =2\left[E_A^iE_B^j\left(n^kn^l-\delta^{kl}\right) \partial_t\Big(
		S_{ij|kl}-S_{kj|il}-S_{il|kj}+S_{kl|ij}\Big) \right]^{\mathrm{TF}}.
\end{aligned}
\label{AlternateFiveGroupSum}
\end{equation}
Define
\begin{equation*}
	\mathcal R_{ikjl}:= S_{ij|kl}-S_{kj|il}-S_{il|kj}+S_{kl|ij}.
\end{equation*}
The symmetry of the stress moments makes $\mathcal R_{ikjl}$ antisymmetric under
$i\leftrightarrow k$ and under $j\leftrightarrow l$, and symmetric when the two index pairs
are interchanged.  Its fully tangential components are
\begin{equation*}
	\mathcal R_{ACBD}:=E_A^iE_C^kE_B^jE_D^l\mathcal R_{ikjl}.
\end{equation*}
Each antisymmetric pair in two dimensions has only one independent component.  Hence, for
some scalar $F$,
\begin{equation*}
	\mathcal R_{ACBD} =F\epsilon_{AC}\epsilon_{BD}
	=F\left(\gamma_{AB}\gamma_{CD}-\gamma_{AD}\gamma_{CB}\right).
\end{equation*}
Using $n^kn^l-\delta^{kl}=-E_C^kE_D^l\gamma^{CD}$ in (\ref{AlternateFiveGroupSum}) now gives
\begin{equation}
	D_{AB} =-2\left[\gamma^{CD}\partial_t\mathcal R_{ACBD}\right]^{\mathrm{TF}}
	=-2\left[\dot F\gamma_{AB}\right]^{\mathrm{TF}}.
	\label{AlternateFiveGroupTrace}
\end{equation}
The two-dimensional trace-free part of $\dot F\gamma_{AB}$ vanishes, and hence
\begin{equation}
	D_{AB}=0.
\end{equation}
The cancellations work differently from those in the Comp\`ere et al.  truncation. The
conservation equations have been used and there are no additional conditions on the stress
moments. 

It is easy to consider the order-0 ($L_0 = 0$) and order-1 ($L_0 = 1$) truncations. In both
cases the sum of groups of five terms vanishes identically leading to $D_{AB} = 0$.


\begin{thebibliography}{99}
%
	\bibitem{ABK-I} A Ashtekar, B Bonga and A Kesavan, Class. Quant.  Grav.  {\bf 32},
		025004 (2015), [arXiv:1409:3816];
%
	\bibitem{ABK-II} A Ashtekar, B Bonga and A Kesavan, Phys. Rev.  {\bf D92}, 044011
		(2015), [arXiv:1506:0615]; 
%
	\bibitem{ABK} A Ashtekar, B Bonga and A Kesavan, Phys.  Rev.  Lett. {\bf 116},
		051101, (2016), [arXiv:1510.04990];
%
	\bibitem{ABK-III} A Ashtekar, B Bonga and A Kesavan, Phys. Rev.  {\bf D92}, 10432,
		(2015), [arXiv:1510.05593];
%
	\bibitem{NoIncoming} A Ashtekar and S Bahrami, Phys. Rev. {\bf D100}, 024042 (2019),
		[arXiv:1904.02822];
%
	\bibitem{Saw1} V L Saw, Phys. Rev. D {\bf 94}, 104004 (2016), [arXiv:1605.05151];
%
	\bibitem{Saw2} V L Saw, Phys. Rev. D {\bf 97}, 084017 (2018), [arXiv:1711.01808];
%
	\bibitem{CFR} G Comp\`ere, A Fiourucci and R Ruzziconi, Class.  Quant. Grav. {\bf
		36}, 195017, (2019) [arXiv:1905:00971];
%
	\bibitem{VRS} H J de Vega, J  Ramirez and N  Sanchez, Phys. Rev.  {\bf D60}, 044007,
		(1999), [arXiv:astro-ph/9812465].
%
	\bibitem{DH-1} G Date and Sk J Hoque, Phys. Rev. {\bf D94}, 064039 (2016),
		[arXiv:1510.07856];  
%
	\bibitem{CHK} G Comp\`ere, Sk J Hoque and E S Kutluk, Class.  Quantum Grav. {\bf
		41}, 155006, (2024), [arXiv:2309.02081];
%
	\bibitem{NKD80} G Date, Linearized gravitational waves in de Sitter space-time,
		[arxiv:2411.16371]; 
%
	\bibitem{BBP} B Bonga, C Bunster and A P\'erez, Phys.  Rev.  {\bf D108}, 064039,
		(2023), [arXiv:2306.08029];
%
	\bibitem{BCFOS} L Blanchet, G Comp\`ere, G Faye, R Oliveri and A Siraj, 	J.
		High Energ. Phys.  {\bf 2021}, 29 (2021), [arxiv:2011.10000].
%
	\bibitem{Harsh:2024kcl} Harsh, Sk J Hoque, S P Kashyap and A Virmani, Class. Quant.
		Grav. {\bf 41}, 225011 (2024), [arXiv:2405.10777;
		doi:10.1088/1361-6382/ad8437].
%
	\bibitem{Harsh:2026vmh} Harsh, O Shetye and A Virmani, Octupolar Gravitational
		Radiation in de Sitter Spacetime, (2026), [arXiv:2608.10593].
%
%
\end{thebibliography}
\end{document}